\documentclass[iop,apj]{emulateapj}
\usepackage{amsbsy}
\usepackage{amsfonts}
\usepackage{amssymb}
\usepackage{bm}
\usepackage{mathrsfs}
\usepackage{pifont}
\usepackage{stmaryrd}
\usepackage{textcomp,bookmark}
\usepackage{portland,xspace}
\usepackage{amsmath,amsxtra}
\usepackage[OT2,OT1]{fontenc}
\usepackage{graphicx}
\usepackage{hyperref}
\newcommand{\msun}{\ifmmode {{M}_\odot} \else  {${M}_\odot$} \fi}
\newcommand{\mpc}{\ifmmode {{\rm Mpc}} \else  {${\rm Mpc}$} \fi}
\newcommand{\kpc}{\ifmmode {{\rm kpc}} \else  {${\rm kpc}$} \fi}
\newcommand{\hmpc}{\ifmmode {h^{-1}{\rm Mpc}} \else  {$h^{-1}{\rm Mpc}$} \fi}
\newcommand{\hkpc}{\ifmmode {h^{-1}{\rm kpc}} \else  {$h^{-1}{\rm kpc}$} \fi}
\newcommand{\amin}{\ifmmode {{\rm arcmin}} \else  {${\rm arcmin}$} \fi}
\newcommand{\mv}{\ifmmode {{M}_{vir}} \else  {${M}_{vir}$} \fi}
\newcommand{\mt}{\ifmmode {{M}_{200}} \else  {${M}_{200}$} \fi}
\newcommand{\cv}{\ifmmode {{c}_{vir}} \else  {${c}_{vir}$} \fi}
\newcommand{\ct}{\ifmmode {{c}_{200}} \else  {${c}_{200}$} \fi}
\newcommand{\rv}{\ifmmode {{r}_{vir}} \else  {${r}_{vir}$} \fi}
\newcommand{\rt}{\ifmmode {{r}_{200}} \else  {${r}_{200}$} \fi}
\newcommand{\rdlt}{\ifmmode {{r}_{\scriptscriptstyle\Delta}} \else  {${r}_{\scriptscriptstyle\Delta}$} \fi}
\newcommand{\mdlt}{\ifmmode {{M}_{\scriptscriptstyle\Delta}} \else  {${M}_{\scriptscriptstyle\Delta}$} \fi}
\newcommand{\cdlt}{\ifmmode {{c}_{\scriptscriptstyle\Delta}} \else  {${c}_{\scriptscriptstyle\Delta}$} \fi}
\newcommand{\smM}{\ifmmode {\scriptscriptstyle M} \else  {${\scriptscriptstyle M}$} \fi}
\newcommand{\smT}{\ifmmode {\scriptscriptstyle T} \else  {${\scriptscriptstyle T}$} \fi}
\newcommand{\hh}{\ifmmode {h^{-1}} \else  {$h^{-1}$} \fi}
\newcommand{\vth}{\ifmmode {\vec{\theta}} \else  {$\vec{\theta}$} \fi}
\newcommand{\ie}{i.e.,~}
\newcommand{\eg}{e.g.,~}
\newcommand{\etal}{et al.}

\newcommand{\lgg}{\ifmmode {{\rm log}} \else  {${\rm log}$} \fi}
\newcommand{\sn}{\ifmmode {\sigma_n} \else  {$\sigma_n$} \fi}
\newcommand{\snsq}{\ifmmode {\sigma_n^2} \else  {$\sigma_n^2$} \fi}
\begin{document}

\title{Effects of center offset and noise on weak-lensing derived concentration--mass relation of dark matter halos}
\author{Wei Du $\&$ Zuhui Fan}
\affil{Department of Astronomy, School of Physics, Peking University, Beijing 100871, China}
\email{fanzuhui@pku.edu.cn}

\begin{abstract}
With the halo catalog from the {\it Millennium Simulation}, we analyze the weak-lensing measured density profiles for clusters of galaxies, paying attention to the determination of the concentration--mass ($c$--$M$) relation which can be biased by the center offset, selection effect, and shape noise from intrinsic ellipticities of background galaxies.
Several different methods of locating the center of a cluster from weak-lensing effects alone are explored. We find that, for intermediate redshift clusters, the highest peak from our newly proposed two-scale smoothing method applied to the reconstructed convergence field, first with a smoothing scale of $2\arcmin$ and then $0\overset{\prime}{.}5$, corresponds best to the true center.
Assuming the parameterized Navarro--Frenk--White profile, we fit the reduced tangential shear signals around different centers identified by different methods.
It is shown that, for the ensemble median values, a center offset larger than one scale radius $r_s$ can bias the derived mass and
concentration significantly lower than the true values, especially for low-mass halos. However, the existence of noise can compensate for the
offset effect and reduce the systematic bias, although the scatter of mass and concentration becomes considerably larger.
Statistically, the bias effect of center offset on the $c$--$M$ relation is insignificant if an appropriate center finding method is adopted.
On the other hand, noise from intrinsic ellipticities can bias the $c$--$M$ relation derived from a sample of weak-lensing analyzed clusters if a simple
$\chi^2$ fitting method is used. To properly account for the scatter and covariance between $c$ and $M$, we apply a Bayesian method to improve the statistical analysis
of the $c$--$M$ relation. It is shown that this new method allows us to derive the $c$--$M$ relation with significantly reduced biases.
\end{abstract}

\section{Introduction}

In the cold dark matter paradigm, dark matter halos play critical roles in hosting the formation of
luminous objects and shaping the observable universe. The structures of dark matter
halos themselves also carry important cosmological information closely related to properties of dark matter particles,
as well as different astrophysical processes affecting their formation and evolution.
Numerical simulations reveal an approximately universal behavior of the density profile
with $\rho \propto r^{-1}$ at the inner part and $\rho \propto r^{-3}$ at the outer part of halos.
Different fitting models for the density profile have been proposed, such as Navarro--Frenk--White
(NFW; \citealt{1996ApJ...462..563N,1997ApJ...490..493N}), generalized NFW \citep{1990ApJ...356..359H,1996MNRAS.278..488Z} and Einasto profiles \citep{1965TrAlm...5...87E,2012A&A...540A..70R}.
Among others, the profile proposed by \citet{1996ApJ...462..563N,1997ApJ...490..493N}, namely, the NFW profile,
has been widely adopted to describe the density distribution. It can be written as
\begin{equation}\label{eq:rho}
\rho(r)=\frac{\rho_s}{(r/r_s)(1+r/r_s)^2},
\end{equation}
where $\rho_s$ and $r_s$ are the characteristic density and scale of a halo respectively.
Given the halo radius \rdlt, another equivalent set of parameters, the concentration $\cdlt$ and the
mass $\mdlt$, is normally used to characterize the density profile, where $\cdlt=\rdlt/r_s$ and
$\mdlt=(4\pi/3)\Delta\rho_{\rm crit}\rdlt^{3}$. Here $\Delta$ is the average overdensity parameter within
$\rdlt$ with respect to the critical matter density $\rho_{\rm crit}$ of the universe, and $\Delta=200$ or $\Delta=\Delta_{\rm vir}$
for the virialized region of a halo are often adopted in different analyses.

The relation between $\cdlt$ and $\mdlt$ ($c$--$M$ relation) is related to the formation history of dark matter halos and has been studied extensively with simulations \citep[e.g.,][]{2001MNRAS.321..559B,2008MNRAS.390L..64D,2008MNRAS.387..536G,2009ApJ...707..354Z,2012MNRAS.423.3018P,2014MNRAS.441..378L}. At low redshifts, the relation of the two can
be approximately described by a power law given by
\begin{equation}\label{eq:cm}
\lgg\,\cdlt=\lgg\,A(z)+\alpha(z) \lgg\,\frac{\mdlt}{M_p}
\end{equation}
in log-space where both $A(z)$ and $|\alpha(z)|$ are decreasing functions of redshift \citep{2008MNRAS.387..536G,2011MNRAS.411..584M}.
With certain variations for the results obtained from one group to another, simulation studies find the slope
$\alpha\sim -0.1$ at redshift $z\sim 0$.
A is more sensitive to cosmological models than $\alpha$ and depends on the definition of the halo mass
(such as $\Delta=200$ or $\Delta_{\rm vir}$) and choice of the pivot mass $M_p$
\citep[e.g.,][]{2004A&A...416..853D,2006ApJ...646..815S,2008MNRAS.390L..64D,2011MNRAS.411..584M,2012MNRAS.424.1244F,2013ApJ...766...32B}.

Because of its cosmological significance, extensive efforts have been made
to observationally measure the density profile of dark matter halos and further investigate their $c$--$M$ relation. For that,
clusters of galaxies are the most important targets. They are known as the largest virialized objects in the Universe, and
their formation and evolution are dominantly affected by gravitational processes. Among others, gravitational lensing analyses provide a direct way to study the dark matter distribution of clusters of galaxies
\citep[e.g.,][]{2001PhR...340..291B,2002MNRAS.337.1269W,2012ApJ...748...56S,2011A&ARv..19...47K, 2013SSRv..177...75H}.

Concerning the $c$--$M$ relation from lensing studies, \citeauthor{2010PASJ...62..811O} (2010, hereafter Ok10) perform detailed weak-lensing analyses for $30$ X-ray selected clusters and
show that the power index of the $c$--$M$ relation is $\alpha\sim -0.4$, significantly steeper than the simulation result.
\citeauthor{2012MNRAS.420.3213O} (2012, hereafter Og12) carry combined strong and weak-lensing studies for $28$ strong-lensing selected clusters and obtain a slope of $\alpha\sim -0.59$,
noting that the expected $\alpha$ is $\sim-0.2$  by taking into account the strong-lensing selection bias.
Combining the richness measurements and the Einstein radii from strong arcs, \citet{2012ApJ...761....1W} study $10$ strong-lensing clusters with redshift
$0.26\le z\le 0.56$. They find $\alpha\sim -0.45$ and show that the steep slope is mainly driven by the low-mass clusters in their sample.
\cite{2013MNRAS.434..878S} compile the weak-lensing analyses of $31$ massive clusters at high redshift with $z\sim (0.8, 1.5)$, and also find
a steep $\alpha\sim -0.8$.
It has been suggested that the steep slope from observations may indicate that physical effects, such as baryon cooling and
the dynamical status of clusters, play non-negligible roles in affecting the mass distribution of dark matter halos
\citep[e.g.,][]{2008JCAP...08..006M,2010MNRAS.405.2078M,2010A&A...524A..68E,2012MNRAS.420.3213O,2012MNRAS.424.1244F,2013MNRAS.434..878S}.

On the other hand, \citet{2007MNRAS.379..190C} combine a variety of observational results with their
$10$ strong lensing clusters and find a slope of $\alpha\sim -0.14$, consistent with simulation results.
Based on data from the Sloan Digital Sky Survey (SDSS; \citealt{2000AJ....120.1579Y}),
\citet{2008JCAP...08..006M} perform stacked weak-lensing analyses around galaxies binned by their luminosities and MaxBCG clusters binned by their richness. The mass range covers $\sim 10^{12}\msun$ to $\sim 10^{15}\msun$.
They also show that their results are consistent with the $c$--$M$ relation with the slope parameter $\alpha\sim -0.13$.

Noticing considerable differences from different analyses, in order to compare
with simulation results and draw conclusions properly, it is therefore necessary to understand how various effects
can influence the lensing derived density profile and mass, and further the $c$--$M$ relation of dark matter halos.
Focusing on weak-lensing studies, in this paper, we specifically investigate how the $c$--$M$ relation is affected by center choice of clusters and shape noise of background galaxies using the halo catalog from the {\it Millennium Simulation} (MS; \citealt{2005Natur.435..629S}).

The center identification for a cluster of galaxies is important in weak-lensing analyses. Optical or X-ray observations
are commonly used for locating the center defined as the position of the brightest cluster galaxy (BCG)
or the X-ray centroid \citep[e.g.,][]{2007ApJ...660..221K,2007ApJ...660..239K,2010A&A...524A..68E,2012ApJ...757....2G}. Regardless of the very possible offsets between these observables and true halo centers \citep{2006A&A...451..395C,2012MNRAS.426.2944Z,2012MNRAS.420.2120M}, misidentification of, e.g., BCG, can
result in a large fraction of offcenters from true ones \citep[e.g.,][]{2007arXiv0709.1159J,2012MNRAS.426.2944Z}.
On the other hand, for massive clusters with high-quality, weak-lensing observations, it is possible to
find the centers of individual clusters self-consistently using lensing data alone \citep[e.g.,][]{2010MNRAS.405.2215O, 2010A&A...520A..58I, 2012A&A...546A..79I}.
Expectedly, tracing the potential center of a cluster is the ideal case; however, in reality, the weak-lensing determined center
can be offset from the true center due to various effects \citep[e.g.,][]{2006A&A...451..395C, 2010ApJ...719.1408F}. With simulated clusters,
\cite{2012MNRAS.419.3547D} demonstrate that the noise from intrinsic ellipticities of source galaxies and the smoothing procedures
dominantly lead to the offsets.

The general conclusion is that the center offset can affect the determination of the density profile, particularly on small scales. For example, the halo mass
derived from the stacked lensing signals can be systematically low if the center offset effects are not properly considered
\citep[e.g.,][]{2010MNRAS.405.2078M, 2012ApJ...757....2G}. In this paper,
we systematically explore four different methods of locating weak-lensing centers, one for shear field and three for convergence maps, using
the simulated clusters from MS. Their influences on the weak-lensing-derived $c$--$M$ relation are statistically evaluated .

In addition, we also pay particular attention to the effects of shape noise from intrinsic ellipticities
emphasizing their impacts on the weak-lensing-derived $c$--$M$ relation. Significant studies have been done to understand the uncertainties
in weak-lensing determination of cluster mass and profile due to different effects,
including the nonsphericities of cluster mass distributions and the substructures therein, the projection effects of correlated and
uncorrelated large-scale structures, profile fitting methods, the shape noise of source galaxies, etc.
\citep[e.g.,][]{2007MNRAS.380..149C, 2010MNRAS.405.2078M, 2011MNRAS.412.2095H, 2011MNRAS.414.1851O, 2011ApJ...740...25B, 2012MNRAS.421.1073B, 2012MNRAS.426.1558G, 2014MNRAS.440.1899G}. Most of these studies concentrate on the
bias and scatters themselves. Some of them discuss the consequent impact on weak-lensing-derived $c$--$M$ relation
without systematically analyzing how the impact depends on the level of scatter and their correlation \citep{2012MNRAS.421.1073B, 2012MNRAS.426.1558G, 2014MNRAS.440.1899G}.

In our study, we perform systematic analyses about the impact of the shape noise on the $c$--$M$ relation. By taking into account the scatter in concentration and mass and
covariance between these two variables, we use a Bayesian method that includes selection effect based on halo mass function to infer the realistic $c$--$M$ relation from weak-lensing observed $(c,\,M)$. To account for the sample variance, we also carry out Monte Carlo analyses with respect to $19$ clusters of Ok10 and $28$ clusters of Og12 and compare with their observational results on the $c$--$M$ relation.

The rest of the paper is organized as follows. We present the mock data generation for weak-lensing analyses in Section 2.
Different center identification methods and profile fitting are described in Section 3.
Section 4 contains the detailed results. Discussions are given in Section 5.

\section{Mock data construction}

The weak-lensing effects from a single cluster can be described by the Jacobian matrix \citep[e.g.,][]{2001PhR...340..291B}
\begin{equation}\label{eq:jac}
\mathcal{A}=
\left(
              \begin{array}{cc}
                1-\kappa-\gamma_1 & -\gamma_2 \\
                -\gamma_2 & 1-\kappa+\gamma_1 \\
              \end{array}
            \right),
\end{equation}
where $\gamma_1$ and $\gamma_2$ are the two components of the lensing shear written in the complex form as $\boldsymbol \gamma=\gamma_1+i\gamma_2$,
and $\kappa=\Sigma/\Sigma_{\rm crit}$ is the lensing convergence which is the ratio of the projected mass density $\Sigma$
of the cluster to the critical surface density $\Sigma_{\rm crit}$ defined by
\begin{equation}\label{eq:csd}
\Sigma_{\rm crit}=\frac{c^2}{4\pi G}\frac{D_s}{D_dD_{ds}}
\end{equation}
where $D_s$, $D_d$, and $D_{ds}$ are the angular diameter distances from the observer to the lensed source, to the lens, and
from the lens to the source, respectively. The lensing induced shape distortion for a background source is given by the two eigen values of the
Jacobian matrix. Specifically, the source ellipticity is defined as
\begin{equation}
\boldsymbol \epsilon = \frac{1-b/a}{1+b/a}\exp(2{\rm i}\varphi),
\label{epsilon}
\end{equation}
where $a$ and $b$ are the two axial lengths of the source image obtained from the quadrupole moment
of the light distribution. The observed ellipticity can be written as
\begin{equation}\label{eq:eps}
\boldsymbol \epsilon=\left\{
           \begin{array}{ll}
             \displaystyle\frac{\boldsymbol \epsilon_s+\boldsymbol g}{1+\boldsymbol g^\ast\boldsymbol \epsilon_s} &
\hbox{if $|\boldsymbol g|\le 1$} \\\\
             \displaystyle\frac{1+\boldsymbol g \boldsymbol \epsilon_s^\ast}{\boldsymbol \epsilon_s^\ast+\boldsymbol g^\ast}
& \hbox{if $|\boldsymbol g|\ge 1$}
           \end{array}
         \right.
\end{equation}
where $\boldsymbol \epsilon_s$ is the intrinsic ellipticity of the source galaxy and $\boldsymbol g=\boldsymbol \gamma/(1-\kappa)$
is the reduced lensing shear. By assuming randomly orientated intrinsic ellipticities, an unbiased
estimate of $\boldsymbol g$ (or $1/\boldsymbol g*$) can be obtained by averaging over the observed $\boldsymbol \epsilon$
\citep{1997A&A...318..687S}. Then the cluster mass distribution can be analyzed either through parametric modeling
or no-parametric studies.

In our analyses, we
study the density profile of the simulated weak-lensing clusters, assuming it follows the NFW profile.
We are mainly interested in the radial profiles of clusters and thus use the reduced tangential shears with respect to their
chosen centers as our basic quantities. The tangential component of the observed ellipticity can be
calculated by
\begin{equation}\label{eq:ts}
\epsilon_t=-[\epsilon_1{\rm cos}(2\phi)+\epsilon_2{\rm sin}(2\phi)],
\end{equation}
and the reduced tangential shear at the radial position $r$ can then be obtained by averaging $\epsilon_t$ over the galaxies within a ring around $r$. We can also reconstruct the lensing convergence field $\kappa$ iteratively from $\boldsymbol g$ according to
the relation between $\boldsymbol \gamma$ and $\kappa$ \citep{1993ApJ...404..441K,1995A&A...297..287S,1997A&A...318..687S}.

In our studies, the dark matter halo catalog is built on the {\it Millennium Simulation} \citep{2005Natur.435..629S}, which
follows $2160^3$ dark matter particles in a periodic box of $500h^{-1}\hbox{Mpc}$ assuming a flat $\Lambda$CDM cosmology.
The cosmological parameters are $\Omega_m=\Omega_{dm}+\Omega_b=0.25$, $\Omega_b=0.045$, $\Omega_\Lambda=0.75$,
$h=0.73$, $n_s=1$, and $\sigma_8=0.9$. We extract halos from the snapshot of $z=0$ using the Friends-of-Friends (FoF; \citealt{1985ApJ...292..371D}) algorithm with linking parameter, $b=0.2$ and choose the position of the most bound particle as the true halo center for each halo.

For weak-lensing studies, we assume that all of the halos are located at $z=0.2$ as a default, the typical redshift of the weak-lensing observed clusters for source galaxies at $z_s\sim 1$. Because halo structures evolve little from $z=0.2$ to $z=0$, using $z=0$ halos and putting them artificially
at $z=0.2$ should not lead to significantly different results from using halos extracted directly from the snapshot
at $z=0.2$ of the simulation.
It is noted that, while the halos are initially identified with the FoF algorithm, in our weak-lensing analyses,
the mass of each halo is defined by the spherical overdensity method around its halo center.
We adopt the overdensity parameter $\Delta=200$ and define the radius $r_{200}$ within which the average density is $200\rho_{\rm crit}(z=0.2)$.
Correspondingly, the mass, $M_{200}$, is defined as the mass contained within $r_{200}$.
We then include all of the halos with $M_{200}\ge 10^{14}h^{-1}\msun$ in our lens catalog.
For comparison, we divide halos into low-mass halos with $10^{14}h^{-1}\msun\le M_{200}< 5\times 10^{14}h^{-1}\msun$
and high-mass halos with $M_{200}\ge 5\times 10^{14}h^{-1}\msun$. In total, we have $1756$ halos ($1690$
low-mass halos and $66$ high-mass halos) in our final lens catalog. The median value of $ r_{200}$ is $0.82\hmpc$ and
$1.33\hmpc$, respectively.

To generate mock weak-lensing data, we randomly choose a line-of-sight (LOS) direction for each of the low-mass halos
and calculate the projected surface mass density, $\Sigma$, on regular grids. For high-mass halos,
we generate $20$ different projected maps along $20$ different LOS directions for each halo.
We then have $1690$ lensing maps for low-mass halos and $1320$ maps for high-mass halos. It is noted that
the $20$ projections for high-mass halos are only used in statistically analyzing the center offsets.
For the later studies of the $c$--$M$ relation, we use one LOS projection for each halo, low or high mass, in order to faithfully preserve the mass function of dark matter halos. For the $\Sigma$ calculation, we cut a size $6\times r_{200}$ cubic box
around the true center of each halo with the chosen LOS as the $z$ direction, and compute the projected $\Sigma$ using
all the particles within the box. This box is chosen while trying to include the surrounding matter of dark matter halos for consideration.
On the other hand, our setting for the LOS integration range cannot fully account for the projection effects
of the large scale structures (LSS), which can contribute non-negligible scatter to the
weak-lensing-derived mass and concentration parameter for dark matter halos \citep[e.g.,][]{2003MNRAS.339.1155H,2004PhRvD..70b3008D,2011MNRAS.412.2095H,2011MNRAS.414.1851O,2011ApJ...740...25B,2012MNRAS.421.1073B}.
However, it has been shown that their contributions are subdominant when compared to those from intrinsic irregularities of
the halo mass distribution and the shape noise from background galaxies \citep[e.g.,][]{2011ApJ...740...25B, 2012MNRAS.421.1073B},
and therefore, they are not expected to significantly affect our main results.

For the lens redshift $z_d=0.2$, we then obtain $\Sigma$ maps
with a squared area of $(6r_{200})\times (6r_{200})$, sampled on a regular grid with pixel size $5 h^{-1}\kpc$
($\sim2\overset{''}{.}1$) using triangular shaped cloud algorithm \citep{1981csup.book.....H}. For background galaxies, we assume they are all at $z_s=1$.
We then have $\Sigma_{\rm crit}(z_d=0.2;z_s=1)\approx 3.34\times 10^{15}\msun\mpc^{-2}$. The convergence map can thus be obtained by
$\kappa=\Sigma/\Sigma_{\rm crit}$. The corresponding shear $\boldsymbol \gamma$ maps are inverted directly from $\kappa$ maps through
Fourier transformation \citep[e.g.,][]{1993ApJ...404..441K,2004MNRAS.350.1038C,Schneider2006,2012MNRAS.421.1073B}. For weak-lensing analyses from tangential reduced shears,
we populate background galaxies in a field of view (FOV) of $30\times 30 \hbox{ arcmin}^2$ around the true center of each halo, the typical FOV for weak-lensing cluster observations. This corresponds to $5.74\times 5.74\mpc^2$ in physical sizes.
Note that this FOV is generally smaller than that of our $\kappa$ and $\gamma$ maps described above, except for the smallest halo in consideration with an angular scale of $6\times r_{200}$, close to $\sim 30~\amin$ for lens redshift $z_d=0.2$.
Therefore, the boundary effects on the $\gamma$ calculation from $\kappa$ using Fourier transformation should be limited.

Similar to \citet{2012MNRAS.421.1073B}, we take into account the magnification effect in populating background galaxies by

\begin{equation}\label{eq:n_g}
n_{g,{\rm lensed}}=\mu^{-0.5}n_{g,{\rm unlensed}},
\end{equation}
where $\mu=1\slash[(1-\kappa)^2-|\gamma|^2]$ is the lensing magnification, and $n_{g,{\rm lensed}}$ and $n_{g,{\rm unlensed}}$
are the lensed and unlensed galaxy number densities, respectively \citep{Schneider2006,2012MNRAS.421.1073B}.
Thus to generate background galaxies, we first populate them randomly in the FOV of $30\times 30 \hbox{ arcmin}^2$ with a given number density significantly higher than the desired one.
Then, for each galaxy, a probability $p_\mu$ from 0 to 1 is randomly given. Only those galaxies with $p_\mu < \mu^{-0.5}$ at their positions
are kept. This way, we get a background galaxy catalog with their surface number density modulated by the lensing magnification effect, $\mu^{-0.5}$.
From this, galaxies are further selected randomly to get the final galaxy catalog with the desired average number density of $n_g=n_{g,{\rm lensed}}$.
We then assign each galaxy an intrinsic ellipticity,
$\boldsymbol \epsilon_s$, following the probability distribution with random phases and $|\boldsymbol \epsilon_s|\le 1$

\begin{equation}\label{eq:ellip}
    p_{\epsilon_s}(|\boldsymbol \epsilon_s|)=2\pi|\boldsymbol \epsilon_s|\frac{\exp(-|\boldsymbol \epsilon_s|^2/\sigma_{\epsilon_s}^2)}
    {\pi\sigma_{\epsilon_s}^2[1-\exp(-1/\sigma_{\epsilon_s}^2)]},
\end{equation}
where $\sigma_{\epsilon_s}$ is the standard deviation of the total $\boldsymbol \epsilon_s$. The `observed' ellipticity for each galaxy is
then obtained by Equation (\ref{eq:eps}), where the lensing signal at the galaxy position is calculated from the grid values
using a cubic convolution algorithm \citep{1983CGIP...23..258P}.

Finally, for a given set of $n_g$ and $\sigma_{\epsilon_s}$, $1690$ sets of low-mass and $1320$ sets of high-mass mock observational data
are generated. We take $n_g=30 \hbox{ arcmin}^{-2}$ and $\sigma_{\epsilon_s}=0.4$
as the default case. Note that in \cite{2012MNRAS.421.1073B}, they set the standard deviation per component to be $0.2$ as their default case.
This corresponds to $\sigma_{\epsilon_s}=\sqrt{2}\times 0.2\approx 0.28$ in our notation, which is smaller than our default case with
$\sigma_{\epsilon_s}=0.4$.

In order to systematically study the noise effects on the weak-lensing-derived $c$--$M$ relation, we also generate
large source galaxy samples with $n_g=80 \hbox{ arcmin}^{-2}$ for each considered halo. To cover a wide range of noise levels,
we consider different $\sigma_{\epsilon_s}$ with $\sigma_{\epsilon_s}=\{0.2, 0.3, 0.4, 0.5\}$.
Then, for each halo, we have four large samples of source galaxies from which we can construct different subsamples with
different $n_g$ and $\sigma_{\epsilon_s}$, and therefore different noise levels.
Observationally, the typical $\sigma_{\epsilon_s}\sim 0.4$ with certain variations, depending on specific observations \citep[e.g.,][]{2000ApJ...532...88H, 2012MNRAS.420.1384S}.
With the sole purpose of demonstrating the systematical trend of the noise effects, the range of $\sigma_{\epsilon_s}$ considered here is a little stretched.
It is also noted that, as we show later, given a smoothing kernel, the noise effect is largely characterized by $\sigma_n=\sigma_{\epsilon_s}/\sqrt{n_g}$,
the dispersion of the mean signal per unit area. Thus, there are some redundancies in our analyses
for different combinations of $(\sigma_{\epsilon_s},n_g)$. It should also be pointed out that, from an observational point of view,
the values of $\sigma_{\epsilon_s}$ and $n_g$ are usually correlated. Good observational conditions result in smaller $\sigma_{\epsilon_s}$
and larger $n_g$, and vice versa. Therefore, it is emphasized again that our analyses here combining different $(\sigma_{\epsilon_s},n_g)$
are solely for systematically demonstrating different noise levels but not for showing the combinations in real observations.
On the other hand, the wide range of $\sigma_n$ in our analyses indeed covers the noise levels from different observations.
For comparison, we also generate two noiseless catalogs with $n_g=30 \hbox{ arcmin}^{-2}$ and an unrealistic $n_g=300 \hbox{ arcmin}^{-2}$,
assuming no intrinsic ellipticities for background galaxies.

For testing the generality of our methods to locate the center, and making comparisons
with the observational result of Og12, we also generate a mock sample by placing the above selected $1690+66$ halos
at $z=0.46$, with one LOS for each halo. We use $\rho_{\rm crit}(z=0.46)$ and $\Delta=200$ to define $r_{200}$ and $M_{200}$.
The FOV is taken to be $16\times 16 \hbox{ arcmin}^2$ by default. In this case, the source redshift is set to be $z_s=1.12$, in accord with that in Og12.

\section{Data analysis}

To constrain the density profile of a dark matter halo with weak-lensing analyses, it is
important to first identify the center of the halo. Here, we explore different center identification methods based
on the weak-lensing `data' alone and further quantify statistically their corresponding center offsets, $R_{\rm off}$,
with respect to the true centers of the halos. We also describe the procedures for the NFW profile fitting from the reduced tangential shear
signals.

\subsection{Center Identification in the Shear Map}

As described in the previous section, we construct mock weak-lensing data for each of the halos in the catalog.
The data contains the positions and `observed' ellipticities for the background galaxies. They are named as shear maps.
In total, for a given set of $n_g$ and $\sigma_{\epsilon_s}$, we have $1690$ shear maps for the low-mass halos using one random LOS direction for each halo and $1320$ maps for the $66$ high-mass halos with $20$ different LOS for each halo.

We first explore the center identification using the $S$-statistic method \citep{1996MNRAS.283..837S}.
The $S$ value at $\boldsymbol \theta_c$ is defined as \citep[e.g.,][]{2010A&A...520A..58I,2012A&A...546A..79I}

\begin{equation}\label{eq:SS}
S(\boldsymbol \theta_c;\theta_{\rm out})=\frac{\sqrt{2}}{\sigma_\epsilon}\frac{\sum_i\epsilon_{t,i}Q_i(x)}{\sqrt{\sum_iQ_i^2(x)}},
\end{equation}
where $\epsilon_{t,i}$ is the tangential component of $\boldsymbol \epsilon_i$ of the galaxy at $\boldsymbol \theta_i$ (see Eq.(\ref{eq:ts})),
$x=|\boldsymbol \theta_i-\boldsymbol \theta_c|/\theta_{\rm out}$ and
\begin{equation}\label{eq:Qx}
Q(x)=\frac{x_c{\rm tanh}(x/x_c)}{x(1+e^{a+bx}+e^{c+dx})}
\end{equation}
with $a=6, b=-150, c=-47, d=50$ and $x_c=0.15$. The filter function, Q, is approximated to
optimize the tangential shear profile corresponding to an NFW density profile \citep{2007A&A...462..875S}. At each $\boldsymbol \theta_c$, we vary $\theta_{\rm out}$
to find the best value to maximize $S$. This maximized
$S$ is taken to the $S$ value at $\boldsymbol \theta_c$. It can be seen that the best $\theta_{\rm out}$ can be
different at different $\boldsymbol \theta_c$. With the $S$ field over the whole map, the position with the highest $S$ value
is then chosen as the center of the halo \citep{2007A&A...462..875S}.

Specifically, for each shear map, we first divide it into $20^2$ regular grids, and calculate $S$ on each of
the grid points. We then locate the position of the highest $S$ value among the $20^2$ grids. Around this peak position,
we consider a smaller region with a size of two grids along each direction and further divide it into $20^2$ finer grids.
The same analysis as the last step is done to find the highest peak on the finer grids. Around this new peak position,
we perform the analyses another time with even finer grids to find the final highest shear peak (SP). This is defined as
the center of the halo. The final grid size for FOV of $30\times 30 \hbox{ arcmin}^2$ is $[(1800''/20\times 2)/20\times 2]/20\sim 1''$. The $\theta_{\rm out}$ is restricted in the range of $[2,20]~\amin$ and $[1,12]~\amin$ for $z=0.2$ and $z=0.46$ catalogs, respectively.

To test if the use of grids in our searching method can affect the center finding, we also
run the nongrid method of Levenberg--Marquardt \citep{Levenberg1944,Marquardt1963,2009ASPC..411..251M}.
We find that, for the case without noise from intrinsic ellipticities of source galaxies, the results of identified centers are nearly the same
as those of our grid-based search, for both low- and high-mass halos. With the noise included, the results from the two searching methods
are also largely consistent with each other for high-mass halos. However, for low-mass halos, the noise effects are significant
leading to considerable false peaks. These local false maxima severely affect the performance of the Levenberg--Marquardt method.
The results of identified centers are very sensitive to the initial guess of the peak position. With a reasonable setting
of the initial guess not very close to the true center, e.g., taking the initial position at $(5\arcmin, 5\arcmin)$,
the offsets between the Levenberg--Marquardt identified centers and the true ones are rather large.

We therefore use the grid-based searching method in our center finding analyses
both from shear maps here and from convergence maps discussed in Section 3.2.

\subsection{Center Identifications in the Reconstructed Convergence Map}

While lensing shear signals can be extracted directly from observations, the convergence $\kappa$ field represents the
project mass distribution of a halo, and therefore can be used in a more visually clear way to identify the
center of the halo. We also investigate different ways of center finding from the $\kappa$ map.

The reconstruction of $\kappa$ is proceeded following the nonlinear reconstruction method of \citet{1995A&A...297..287S,1997A&A...318..687S}, in which the
$\kappa$ field is obtained iteratively from the estimated reduced shear $\boldsymbol g$ through the integration relation between
$\kappa$ and $\boldsymbol \gamma$ \citep{1993ApJ...404..441K,2000MNRAS.313..524V}. To suppress the noise from intrinsic ellipticities of background galaxies,
we first estimate the smoothed reduced shear signal by

\begin{equation}\label{eq:smg}
g_{sm}(\boldsymbol\theta)=\frac{1}{n_g}\sum_{i}W(|\boldsymbol\theta-\boldsymbol\theta_i|)\boldsymbol\epsilon_i
\end{equation}
where the smoothing function is taken to be Gaussian given by
\begin{equation}\label{eq:wf}
 W(|\boldsymbol\theta-\boldsymbol\theta_i|)=\frac{1}{\pi\theta_G^2}{\rm exp}(-\frac{|\boldsymbol\theta-\boldsymbol\theta_i|^2}{\theta_G^2}).
\end{equation}
Here, $\theta_G$ is the smoothing scale.
As demonstrated in \cite{2004MNRAS.350..893H}, for maximizing the signal-to-noise ratio of an NFW cluster, the optimal choice of the Gaussian smoothing scale is
$\theta_G\sim \theta_s=r_s/D_d$. For a typical cluster with $M\ge 1\times 10^{14}h^{-1}\msun$ and at intermediate redshift,
$\theta_s\sim 1\arcmin$. We therefore consider two smoothing scales here with
$\theta_G=0\overset{\prime}{.}5$ and $\theta_G=2\arcmin$, respectively, to further show the effects of different smoothings.

\begin{figure*}
  \centering
  \includegraphics[width=\textwidth]{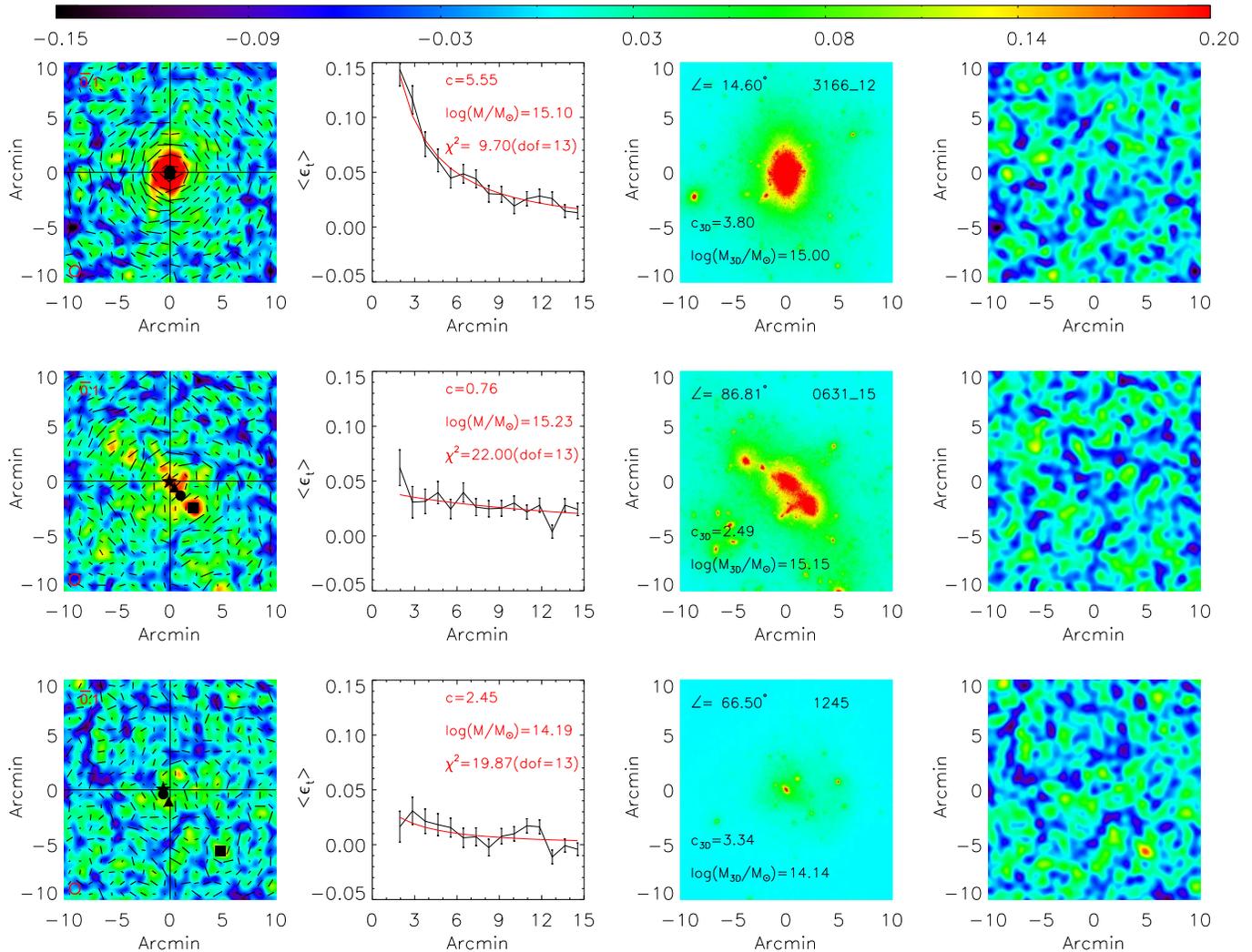}\\
  \caption{Three examples from our mock catalog with lens redshift $z_d=0.2$. The default case with $n_g=30$ and $\sigma_{\epsilon_s}=0.4$ is adopted. The color bar on the top shows the density level in convergence maps. From left to right, the first column shows the reconstructed convergence map in color and the shear field by black sticks with the smoothing scale, $\theta_G=0.5~\amin$. The radius of the red circle at the lower-left corner indicates the smoothing scale. The red stick at the top-left corner denotes the shear value of $0.1$. The filled black squares, circles, five-pointed stars, and triangles show the position of candidate centers found by 05KK, 2KK, two-scale smoothing method 2K05K, and highest shear peak SP, respectively. The second column is the NFW fitting to reduced tangential shears in $15$ equal bins in radius. The error bar in each bin shows the dispersion of the mean reduced tangential shear. The red lines are the best-fit results. The best fitting c and M are shown in the top-right corner with the minimum chi-square. The number of degrees of freedom is 13 for the $\chi^2$ fitting in this column. The third column is the original projected 2D surface density without smoothing. The projected angle with respect to the major axis of the halo and their 3D fitted $c$ and $M$ are presented. The top-right numbers mark the halo IDs in the catalog. The last column is the reconstructed convergence map for the pure noise corresponding to the first column. Note that only the central region of the FOV is shown with $20\times 20~\amin^2$.}\label{Fig:cenfinder}
\end{figure*}

To identify the center of a halo from the weak-lensing $\kappa$ map, the highest peak is the natural choice. We first use the default case with $z_d=0.2$, $n_g=30 \hbox{ arcmin}^{-2}$, and $\sigma_{\epsilon_s}=0.4$ as a test. For $\theta_G=0\overset{\prime}{.}5$, the smoothed reduced shear field and the reconstructed $\kappa$
map are initially sampled on $512^2$ grids in the FOV. For $\theta_G=2\arcmin$, the number of grids is $128^2$.
We first identify the
grid point with the highest $\kappa$ value, excluding regions less than $0\overset{\prime}{.}5$ to the
boundary of each side to eliminate the boundary effects. We denote this grid point as $\boldsymbol x_{K,0}$.
We then consider a squared region of $3\times 3\hbox{ arcmin}^2$ around $\boldsymbol x_{K,0}$ and reconstruct the $\kappa$ field within this
small region from the reduced shear field over the full FOV on finer grids of $180^2$. Within a squared region of $2\times 2\hbox{ arcmin}^2$ around $\boldsymbol x_{K,0}$, we search for the grid point corresponding to the highest $\kappa$ value
on the finer grids, which is finally identified as the center of the map. The second step with finer
grids is to avoid possible miscentering due to the original grid sampling.
Note that the smoothing scale in the second step is the same as first step, \ie $\theta_G=0\overset{\prime}{.}5$ and $2\arcmin$, respectively.
The corresponding centers identified with the above method are labeled as 05KK and 2KK.

In the above analyses, we use a single smoothing scale to smooth the reduced shear field,
either $\theta_G=0\overset{\prime}{.}5$ or $\theta_G=2\arcmin$.
To take the advantage of large smoothing to suppress noise and that of small smoothing to maintain a high resolution,
we also consider a two-scale smoothing method. Specifically, we first choose $\boldsymbol x_{K,0}$ from
the reconstructed $\kappa$ map ($128^2$ grids) smoothed with $\theta_G=2\arcmin$ as the
initial center. We then reconstruct a smoothed $\kappa$ field with $\theta_G=0\overset{\prime}{.}5$ in a squared region
of $3\times 3 \hbox{ arcmin}^2$ sampled on $180^2$ grids around the initial center. The highest peak within the inner
$2\times 2 \hbox{ arcmin}^2$ region of the high-resolution area is finally identified as the center of the map. These centers are labeled as 2K05K.

For the cluster catalog at $z_d=0.46$, the FOV adopted here is $16\times 16\hbox{ arcmin}^2$ for each cluster.
The center identification procedures are the same as described above, except that
the reconstructed  convergence field for each cluster is initially sampled on
$256\times 256$ and $128\times 128$ grid points for $\theta_G=0\overset{\prime}{.}5$ and $\theta_G=2\arcmin$, respectively.

In total, we have three center identifications from $K$ maps and one from reduced shear maps. They are 05KK, 2KK, 2K05K, and SP, respectively.
In Section 4.1, we will show the statistical distributions of the center offset for different identification methods and their impacts on the derived $c$--$M$ relation from a large sample of weak-lensing analyzed clusters.

\subsection{NFW Fitting}

In this paper, we adopt the NFW profile for the density distribution of dark matter halos.
We first fit the 3D NFW profile for each halo with Eq.(\ref{eq:rho}), following the fitting procedure described by \citet{2012MNRAS.421.1073B}.
A halo is divided into $32$ radial bins in logarithmic scale in the range $-2.5\le \log_{10}(r/\rt)\le 0$,
and the density is spherically averaged in
each bin \citep{2008MNRAS.387..536G,2007MNRAS.381.1450N}. Only the bins with $r_{\rm bin}>0.02\rt$ are considered in the fitting. The fitted mass and
concentration parameter are denoted as $M_{\rm 3D}$ and $c_{\rm 3D}$, respectively.

For weak-lensing analyses, we fit the averaged $\langle\epsilon_t\rangle$ to $g_t=\gamma_t/(1-\kappa)$ from the NFW model to derive constraints on the mass and concentration parameter of halos. To calculate the theoretical $\gamma_t$ in each bin, we use the relation
$\gamma_t(R)=\bar \kappa(R) -\kappa(R)$, where $\bar \kappa(R)$ is the mean $\kappa$ over the region within the projected radius $R$ and $\kappa(R)$ is the convergence at $R$. For NFW halos, $\kappa_{\rm NFW}(R)=\Sigma_{\rm NFW}(R)/\Sigma_{\rm crit}$ and the projected surface density $\Sigma_{\rm NFW}$ is calculated by

\begin{equation}\label{eq:sigr}
\Sigma_{\rm NFW}(R)=\int_{-\infty}^{\infty}\rho(r)dz
\end{equation}
where $\rho(r)$ is given by Eq.(\ref{eq:rho}) and $r=\sqrt{R^2+z^2}$ \citep{2000ApJ...534...34W}.
Taking into account the LOS integration length for our mock clusters, we also perform a test for each NFW halo by
using $\pm 3 r_{200}$ as the upper and lower integration limits in Eq.(\ref{eq:sigr}) in the fitting.
The results are very similar to that with the LOS integration set to $\pm \infty$.
Therefore, we use Eq.(\ref{eq:sigr}) in our fitting consistently throughout the paper,
which is also the popularly adopted one in observational analyses.

In the fitting, we use the mass $M$ and the concentration parameter $c$ as the two free parameters instead of $\rho_s$ and $r_s$.
The two sets of parameters are related through $r_s=[3M/(4\pi\Delta \rho_{\rm crit})]^{1/3}/c$ and $\rho_s=\delta_c\rho_{\rm crit}$ with $\delta_c=(\Delta/3)\{c^3/[\ln (1+c)-c/(1+c)]\}$, where we take $\Delta=200$.

To estimate the signals for each weak-lensing map, we first choose a center identified with the methods described above.
Around the center, we divide the region of $[1\arcmin, 15\arcmin]$ into $15$ equal bins for $z_d=0.2$ lenses.
For lens at $z_d=0.46$,  we consider $10$ equal bins in the radial range of $[0\overset{\prime}{.}4, 8\arcmin]$.
We exclude the inner most region from our weak-lensing analyses to avoid the strong lensing effect.
Within the bin $i$, the tangential $\langle\epsilon_t\rangle_i$ is estimated by averaging
$\epsilon_t$ over all of the galaxies within the bin \citep{Schneider2006,2012A&A...546A..79I}. The error is calculated by $\sigma_i=\sigma_{t,i}/\sqrt{N_i}$,
where $\sigma_{t,i}$ and $N_i$ are the standard deviation of the tangential component of galaxy ellipticities
and the number of galaxies within the bin \citep{2000A&A...353...41S}. The weak-lensing constraints on the mass and
concentration of halos are derived by the $\chi^2$ fitting defined as

\begin{equation}
\chi^2=\sum_i\frac{(\langle\epsilon_t\rangle_i-g_{t,{\rm NFW}})^2}{\sigma_i^2}.
\label{chi}
\end{equation}
The Levenberg--Marquardt minimization algorithm \citep{Levenberg1944,Marquardt1963,2009ASPC..411..251M} is used here and also for other $\chi^2$ fittings in this paper.
As a default, uniform priors with $c>0$ and $M>0$ and without specific upper bounds are applied in the NFW fitting.

For the default case, with lens redshift $z_d=0.2$, $n_g=30\hbox{ arcmin}^{-2}$ and $\sigma_{\epsilon_s}=0.4$,
we consider the true center and all of the centers identified by the four different methods.
For other cases with different noise, we use the true center in the analyses.
To see the pure projection effect, we take the case with $n_g=30\hbox{ arcmin}^{-2}$ and $|\boldsymbol \epsilon_s|=0$
and perform the $\chi^2$ fitting analyses.
We use $(M_{\rm 2D},c_{\rm 2D})$, derived from this noise-free case using the true center,
as our reference point. In this noise-free case, $\sigma_i$ is estimated similarly
by $\sigma_i=\sigma_{t,i}/\sqrt{N_i}$ from galaxies' ellipticities (without intrinsic ones) in the bin.
It is noted that although $\sigma_i$ is significantly smaller than that in the noise case, it is not zero
due to the existence of substructures and nonsphericities of dark matter halos.
The case with $n_g=300\hbox{ arcmin}^{-2}$ and $|\boldsymbol \epsilon_s|=0$ is also done and the results are almost the same as
that of $n_g=30\hbox{ arcmin}^{-2}$.

\begin{figure*}
  \centering
  \includegraphics[width=0.8\textwidth]{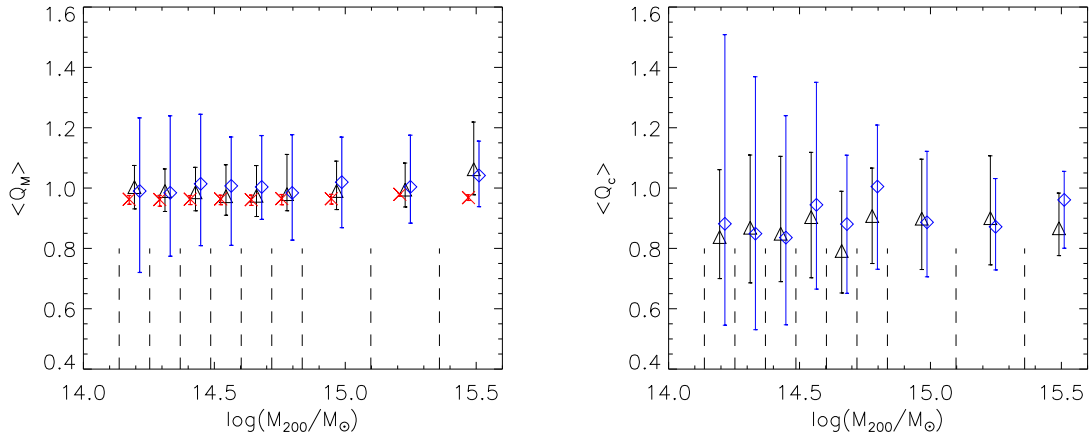}\\
  \caption{Left: median mass normalized by the true mass $\mt$ as a function of $\mt$. The red cross symbols and black triangles are
for $M_{\rm 3D}/M_{200}$ and $M_{\rm 2D}/M_{200}$, respectively. Right: median concentration parameter normalized by $c_{\rm 3D}$. The black triangles are
for $c_{\rm 2D}/c_{\rm 3D}$. Blue diamonds in both panels present the results of default case with $\sigma_\epsilon=0.4$ and $n_g=30 \hbox{ arcmin}^{-2}$
using the true centers. Error bars show the range between the first and third quartiles around the median values. The vertical dashed lines
separate the bins. The horizontal shifts in each bin for different sets of data are for the purpose of clarity. The left six bins
and right three bins are for low- and high-mass halos, respectively.}\label{fig:Qmqc_mvir}
\end{figure*}

\section{Results}

In this paper, we study different center identification methods with weak-lensing analyses alone and quantify the corresponding
offsets. We also systematically investigate the noise effects on weak-lensing-derived $c$--$M$ relation from a sample of clusters.
We emphasize the importance of the Bayesian method in extracting the unbiased $c$--$M$ relation, which properly
takes into account the degeneracy of weak-lensing-determined $(c,M)$.

It is noted that we take $(c_{\rm 2D}, M_{\rm 2D})$ obtained from the noiseless weak-lensing analyses using true centers in the NFW fitting
as our reference point, in which the effects of the intrinsic nonsphericity and substructures of halos are
naturally included \citep[e.g.,][]{2007MNRAS.380..149C,2011ApJ...740...25B,2012MNRAS.421.1073B}. By comparing with the results from the reference case,
we can then systematically study the noise effects and the offcenter effects.

As an illustration, in Figure \ref{Fig:cenfinder}, we show three examples of two high-mass clusters ($3166\_12~\&~0631\_15$) and
one low-mass cluster ($1245$) placed at $z_d=0.2$. Only the central region of $20\times 20 \hbox{ arcmin}^2$
is shown. The first column is the smoothed reduced shear maps (black sticks) overlaid on the
reconstructed convergence maps shown in colors, where the color scale is indicated at the top. The smoothing scale
is $\theta_G=0\overset{\prime}{.}5$ shown as the red circle at the lower left corner of each panel. The red stick at the
upper left corner shows the length-strength scale for the reduced shear, $|g|=0.1$. The longer the sticks, the stronger the shear signals.
Different symbols in each map indicate the center candidates identified with different methods of 05KK, 2KK, 2K05K, and SP.
Each map is centered at the true center of the corresponding halo. The second column shows the profiles of the averaged
tangential $\langle\epsilon_t\rangle$, where the black bars connected by black lines are for the data point, and the red solid lines
are the best NFW fitting results with the two fitting parameters shown in the corresponding panels.
The number of degrees of freedom is $13$ for the $\chi^2$ fitting.
The third column is for the corresponding original $\kappa$ maps of the clusters without smoothing, where the LOS direction
defined by the angle with respect to the major axis and the
label of each halo are shown at the top of each panel. The true 3D concentration and mass of the halos are
shown at the bottom. The last column shows the pure noise reconstructed with $\theta_G=0\overset{\prime}{.}5$ in the $\kappa$ field
resulting from intrinsic ellipticities of background galaxies.

In the first case shown in Figure \ref{Fig:cenfinder} (top panels), the LOS direction is close to the major axis of the halo,
which can be responsible for the high concentration parameter derived from weak-lensing analyses.
For the second halo, the irregularity of the mass distribution is clearly seen. The third halo has a relatively low mass,
and the noise effects are significant. The complex mass distribution and the noise can also lead to large offsets between
our weak-lensing-identified centers and the true center position.

Figure \ref{fig:Qmqc_mvir} shows the statistical comparisons between
$(M_{\rm 3D},c_{\rm 3D})$ and $(M_{\rm 2D},c_{\rm 2D})$ and the $(M,c)$ of the default case with $n_g=30\hbox{ arcmin}^{-2}$ and $\sigma_{\epsilon_s}=0.4$
for the $1756$ simulated halos. The true center of each halo is used in the fitting.
The left panel shows the mass comparison where $Q_M=M/M_{200}$, with $M$ being the value of $M_{\rm 3D}$ (red cross), $M_{\rm 2D}$ (black triangle), or the mass
from the default case (blue diamond). The horizontal axis is $M_{200}$ based on how the halos are binned.
The symbols are for the median values of $Q$, and the error bars show the range of $[1/4,3/4]$ percentiles of the $Q$-value
distributions within different bins. The right panel shows $Q_c=c_{\rm 2D}/c_{\rm 3D}$ (black triangle) and $Q_c=c/c_{\rm 3D}$ with $c$
from the default case (blue diamond). The vertical dashed lines indicate the bin boundaries. Different sets of data points are slightly
shifted horizontally for clarity. It is seen that, when binned based on the true mass of halos $M_{200}$, $M_{\rm 3D}$ is statistically lower than $M_{200}$
by $\sim 4\%$. The bias for $M_{\rm 2D}$ is from $\sim-2\%$ for low-mass to $\sim2\%$ for high-mass halos.
For $c_{\rm 2D}$, it is systematically lower than $c_{\rm 3D}$ by $\sim 13\%$. The results are largely consistent with
previous studies \citep[e.g.,][]{2007MNRAS.380..149C,2011ApJ...740...25B,2012MNRAS.421.1073B, 2012MNRAS.426.1558G}.

The small negative bias of $M_{\rm 3D}$ with respect to $M_{200}$ can be attributed to the deviation of the real density profile
within $r_{200}$ from the NFW profile. We test that both the triaxiality and the substructures contained in a dark matter halo can lead to slight negative biases
in the mass when fitting the spherical NFW profile to its density distribution \citep[e.g.,][]{2012MNRAS.421.1073B}. For the 2D weak-lensing analyses,
it is demonstrated, e.g., \cite{2012MNRAS.421.1073B} and \cite{2012MNRAS.426.1558G}, that the projection along the line of sight makes the impacts
of the complicated mass distribution of halos more significant on the 2D NFW fitting than that on the 3D fitting.
This can explain the relatively large negative bias of $c_{\rm 2D}$ with respect to $c_{\rm 3D}$.

For the default case with noise (blue), the median values of $M$ and $c$ are in good agreement with $M_{\rm 2D}$ and $c_{\rm 2D}$ with little bias.
The scatters are significantly larger as expected. We will see later the scatter in $M$ and $c$ are strongly correlated,
which can lead to a bias in the determination of the $c$--$M$ relation from a sample of weak-lensing-studied clusters
if the binning of clusters is based on their weak-lensing-derived mass.

\subsection{Offsets of Weak-lensing Identified Centers}

Here, we compare different center finding methods, 05KK, 2KK, 2K05K, and SP, by statistically quantifying the corresponding
center offsets with respect to the true centers of halos and their impacts on the weak-lensing determinations of $(c,M)$.

\begin{figure*}
  \centering
  \includegraphics[width=0.8\textwidth]{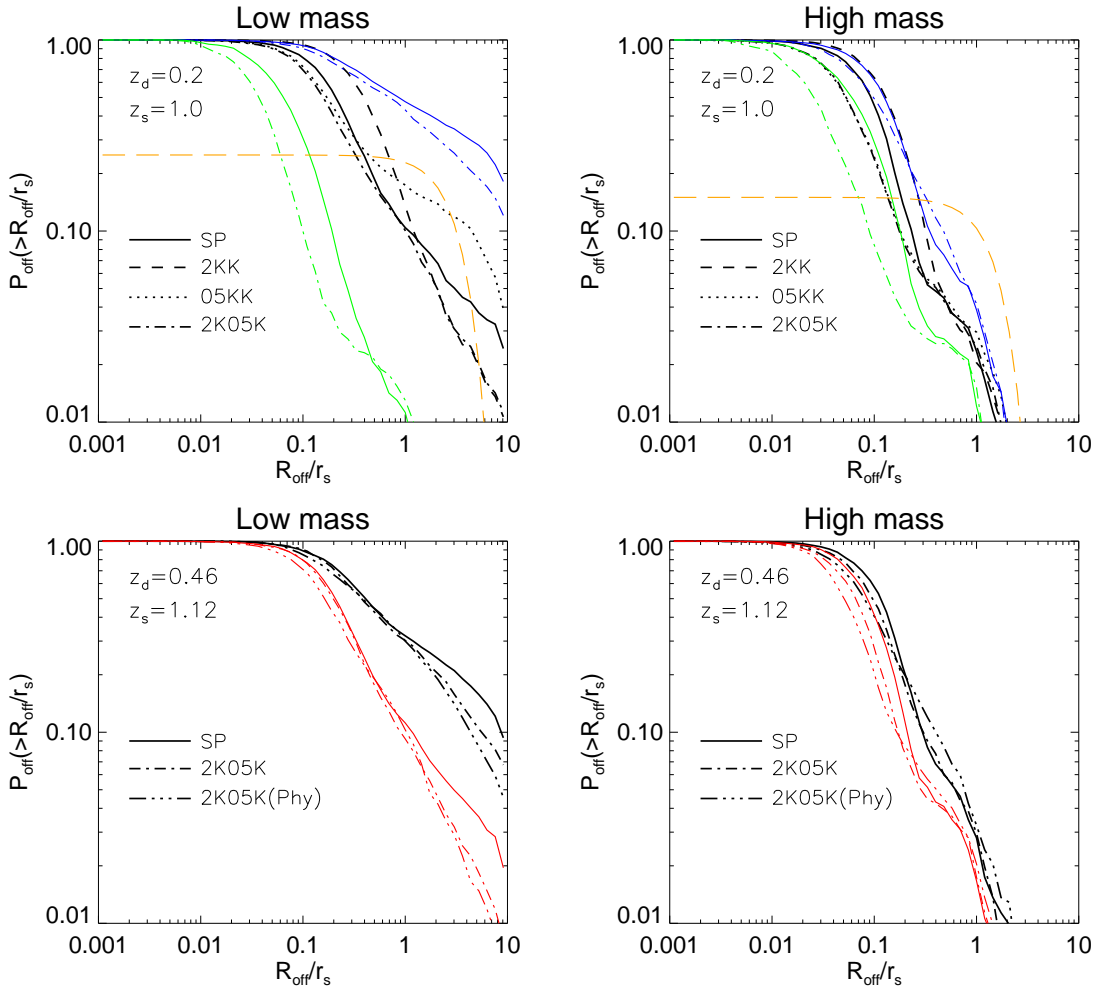}\\
  \caption{Cumulative probability distribution of offset to the true center for the four center candidates identified with different methods. The left and right panels are for low- and high-mass, respectively. Different line styles  correspond to centers from different methods. The green lines show the cumulative distribution for the noise-free case. The blue, black and red are for $n_g=10, 30$ and $64.2$, respectively. The orange long dashed lines are predicted from the offset distribution for misidentified BCGs based on the study of \citet{2007arXiv0709.1159J}.}\label{fig:Peaks_hor}
\end{figure*}

In Figure \ref{fig:Peaks_hor}, we present the cumulative probability distribution, $P_{\rm off}(>R_{\rm off}/r_s)$,
of the center offset, $R_{\rm off}$, for different identification methods. The horizontal axis is the offset $R_{\rm off}$ scaled by $r_s$ for each halo.
The left panels show the results of the $1690$ low-mass halos, where one random LOS is chosen for each halo.
The right panels are for the $66$ high-mass halos where $20$ weak-lensing maps with different LOS are considered for each halo.
The top and bottom panels are for the cases with $z_d=0.2$ and $z_d=0.46$, respectively. The green lines show the cumulative distribution of the noise-free case.
The blue, black and red lines are for $n_g=10, 30$ and $64.2$, respectively, and $\sigma_{\epsilon_s}=0.4$ in all three cases.
Different line styles correspond to the results from different center identification methods as specified in each of the panels.
The orange long dashed lines are the results from \citet{2007arXiv0709.1159J} showing the Gaussian-type  offset distribution for the misidentified BCGs using the
MaxBCG center identification algorithm \citep{2007ApJ...660..221K}. The fraction of misidentified BCGs is estimated to be about $25\%$ and $15\%$ for
the low-mass and high-mass halos, respectively, based on the richness dependence of the misidentified fraction and the richness-mass relation,
and the dispersion of the offset is taken to be $\sigma_s=0.42\hmpc$ \citep{2007arXiv0709.1159J}. We use the median values of $r_s$ with
$r_s=0.18\hmpc$ and $r_s=0.36\hmpc$ for low- and high-mass halos, respectively, to scale the offsets for MaxBCG centers
in plotting the orange lines.

For the default case in the top two panels, we present the results of all four methods.
For 05KK, with a small smoothing scale of $\theta_G=0.5\hbox{ arcmin}$ (black dotted line),
there is a considerable fraction of misidentified centers with offsets to the true centers larger than $r_s$
for low-mass halos, primarily due to the noise effects generating false peaks. The fraction reaches $\sim 20\%$.
For high-mass halos, $\sim 3\%$ with $R_{\rm off}>r_s$ mainly because of the existence of small-scale subclumps.
For 2KK with a relatively large smoothing scale of $\theta_G=2\hbox{ arcmin}$ (black dashed line),
the fraction of large offset due to small-scale structures is reduced considerably. However, this large smoothing
increases the fraction of center offset in the range of $0.1r_s$ to $r_s$. These results are consistent with those of \cite{2012MNRAS.419.3547D}.
For our newly proposed two-scale smoothing methods of 2K05K (black dash dot line), the results are significantly better.
The large-offset fraction is effectively controlled from the first step smoothing with $\theta_G=2\hbox{ arcmin}$.
The second step of high-resolution analyses with $\theta_G=0.5\hbox{ arcmin}$, on the other hand, significantly reduces the smoothing effects
in the offset range of $(0.1,1.0)r_s$.
For SP (black solid line), it involves multiscale and multiresolution analyses.
The results are similar to 2K05K with a somewhat larger fraction with $R_{\rm off}>r_s$.
Overall, the method of 2K05K statistically gives rise to the best center identifications.

For generality, we also consider cases of different noise levels using the two better center identification methods, SP and 2K05K.
For the noise-free case (green lines), both methods perform well and nearly all $R_{\rm off}<r_s$ for both low- and high-mass halos.
For 2K05K, more than $90\%$ of the offsets are smaller than $0.1r_s$.
For the larger noise case with $n_g=10$ and $\sigma_{\epsilon_s}=0.4$ (blue lines), the centers identified by either SP or 2K05K
are not reliable for low-mass halos. For high-mass halos (blue lines in the upper right panel),
the noise effects are relatively weak, and the offset distributions are not significantly different from those of
the default case.

The bottom panels show the application of both 2K05K and SP to the lens cluster catalog at $z_d=0.46$.
The dash dot lines are for 2K05K with $2'$ ($\sim0.5\hmpc$ at $z_d=0.46$) and $0\overset{\prime}{.}5$ ($\sim0.12\hmpc$ at $z_d=0.46$).
Here we also show (dash three dots lines) the results from the two-scale smoothing method where the two smoothing scales are chosen to give rise to the
same physical smoothing scales as those of $\theta_G=2\arcmin$ and $0\overset{\prime}{.}5$ for $z_d=0.2$ lenses. The two physical scales are
$0.28\hmpc$ and $0.07\hmpc$, respectively. It is seen that the results are nearly the same as those shown by dash dotted lines.

The comparisons between the upper and lower panels show that the black (blue) lines in the upper panels are
very similar to the red (black) lines in the lower panels, where the noise levels are very different.
We further find that the performance of the center identification methods discussed here depends on the quantity $E_{\rm SN}$ defined by

\begin{equation}\label{eq:StN}
E_{\rm SN}=\frac{D_{ds}/D_s}{\sigma_{\epsilon_s}/\sqrt{n_g}}\propto\frac{S}{N}.
\end{equation}
It is noted that given a set of clusters, if they are put at different redshifts, the lensing signals are different.
The quantity $E_{\rm SN}$ then reflects more or less the signal-to-noise ratios of clusters at different redshifts, as long as the clusters
do not significantly evolve within the considered redshift range. We have $E_{\rm SN}=10.39$ for the case shown in black in the upper panels
and that shown in red in the lower panels. For the blue lines in the upper and the black lines in the lower panels, $E_{\rm SN}\sim 7$.
Therefore, our analyses show that generally, it is the signal-to-noise level, rather than the noise itself, that determines the performance of
our center identification methods 2K05K (SP). In order to ensure that more than $90\%$ of centers identified by 2K05K have offsets less than $r_s$,
$E_{\rm SN}\sim 10$ or higher is observationally required. In this case, our 2K05K based on weak-lensing analyses alone outperforms that of
BCGs as seen by comparing the orange long-dashed lines in the upper panels representing the offset distribution of misidentified BCGs.

\begin{figure*}
  \centering
  \includegraphics[width=0.8\textwidth]{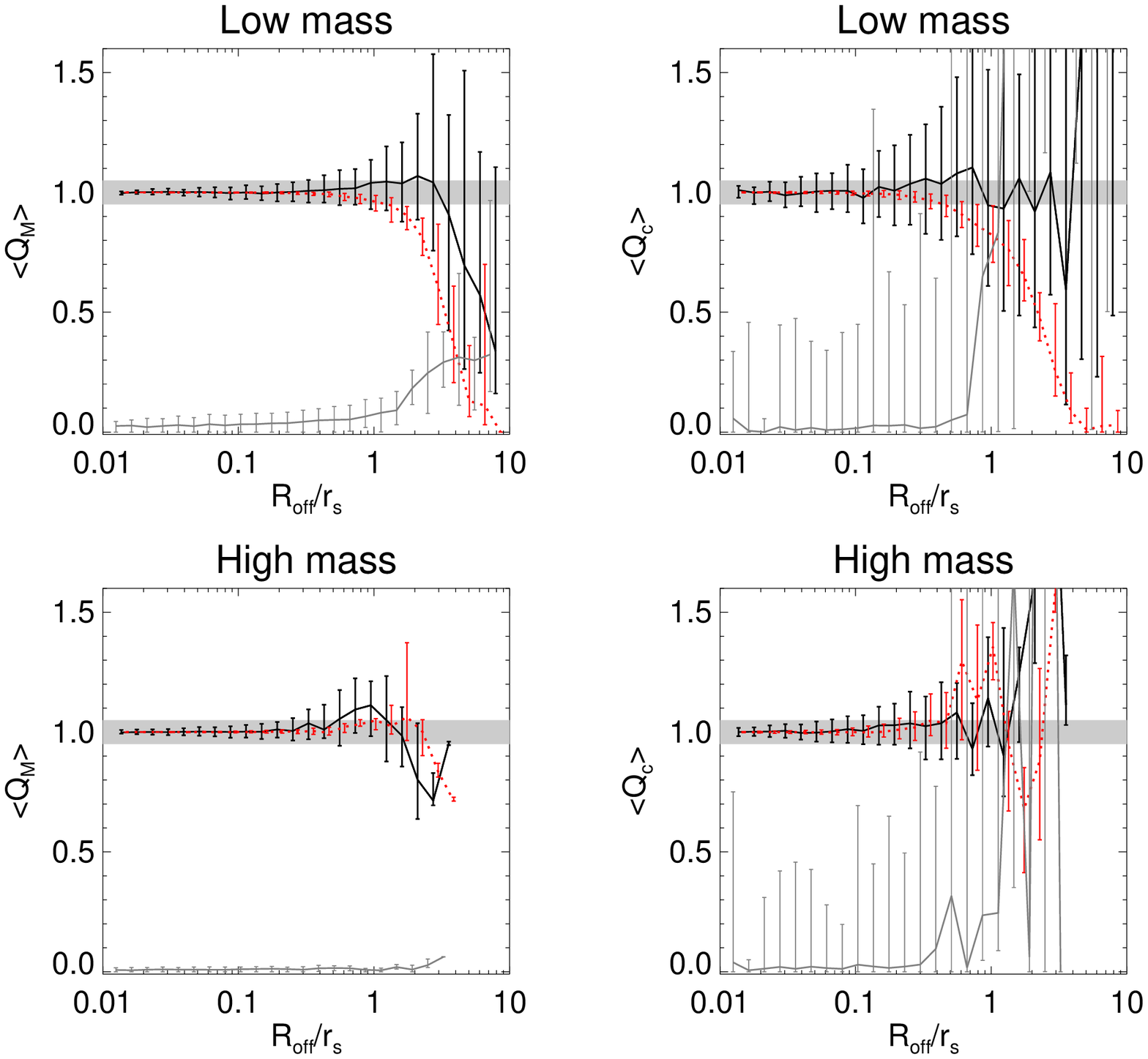}\\
  \caption{Median-normalized mass and concentration parameter as a function of center offset.
The upper and lower panels are for low- and high-mass subsamples, respectively. The left and right panels are for mass and
concentration parameter, respectively. The black and red dotted lines are the results for the default case and noise free case, respectively.
The gray lines are for the results of the pure noise fields. The gray shaded region in each panel marks $5\%$ deviation from unity.
Error bars show the range between first and third quartiles around the median values.}\label{fig:Q_cm}
\end{figure*}

Figure \ref{fig:Q_cm} shows the effects of the center offsets on the weak-lensing determination of $M$ (left) and $c$ (right).
We consider all four centers identified for the default case with $n_g=30\hbox{ arcmin}^{-2}$ and $\sigma_{\epsilon_s}=0.4$.
For the NFW profile fitting, we first analyze the maps with $n_g=30\hbox{ arcmin}^{-2}$ and $|\boldsymbol {\epsilon_s}|=0$, i.e.,
without including the noise in the fitting. This can clearly show us the offcenter effects. The results are
presented in Figure \ref{fig:Q_cm} with red dotted lines where $Q_M$ and $Q_c$ are the normalized mass and concentration, respectively, to the corresponding $M_{\rm 2D}$ and $c_{\rm 2D}$ obtained using the true center for each map. We bin the results by the offset and show the median values with error bars within each bin. The upper and lower panels are for the results of low-mass and high-mass halos, respectively.
The shaded regions indicate $\pm 5\%$ deviation from $1$. We see that for low-mass halos, a large center offset leads to
a systematic bias in both the mass and the concentration parameter, with the latter being more
sensitive to the offset. At the offset of $r_s$,
the lensing-derived mass is lower than $M_{\rm 2D}$ by $\sim 5\%$, and the concentration parameter is lower than $c_{2D}$
by about $25\%$.
It is noted that, for low-mass halos, the intrinsic mass distribution of dark matter halos is relatively simple and regular
without significant large substructures. For high-mass halos, however, the situation is somewhat
complicated because their intrinsic irregularities are relatively high. While the center offset for low-mass halos
is largely induced by noise, the identified center can correspond to true subclumps for high-mass halos
(e.g., see the different identified centers in the middle left panel of Figure \ref{Fig:cenfinder}).
Those subclumps appear more centrally concentrated along specific LOS than their host halos and
therefore become the highest peaks in either shear maps or convergence maps. So, the concentration derived by fitting the signal around those subclumps may be even higher. It is also interesting to note that the mass for high-mass halos is less affected by the offset effect.

We now analyze weak-lensing constraints with noise included. The black solid lines in Figure \ref{fig:Q_cm} show the results for the
default case with $n_g=30\hbox{ arcmin}^{-2}$ and $\sigma_{\epsilon_s}=0.4$. The lines in different panels
have the same meaning as the noise-free case. The quantities $Q_M$ and $Q_c$, derived from weak-lensing analyses
for a map using four different centers, are with respect to $M$ and $c$ derived from the same map around the true center.
Comparing to the results indicated by red dotted lines, we see that while the scatter is significantly larger,
the existence of noise largely reduces the systematic bias for low-mass halos.
That is because for these halos, the large offsets are usually associated with high-noise peaks.
When fitting the signals, including noise, to the NFW profile around these false peaks,
the chance alignments of intrinsic ellipticities leading to the false peaks contribute to the inferred mass with a high concentration.
The median mass and concentration derived from fitting the NFW profile to the corresponding pure noise around the
peaks are shown by gray lines in Figure \ref{fig:Q_cm}. They are normalized by $M$ and $c$ for the default case using the true centers.
It is seen that, for low-mass halos, the noise near the false centers indeed has significant effects on $M$ and $c$,
compensating for the impacts of large center offsets. For high-mass halos, the noise effects are relatively weak,
and the compensating effects are less significant than that for low-mass halos.

\subsection{$c$--$M$ Relation Analyses}

The $c$--$M$ relation analyses discussed in this paper concern a sample of, but not a single, weak-lensing-studied cluster.
Specifically, for a cluster $i$ in a given sample containing $N$ clusters, by fitting the tangential reduced shear signals to that predicted by the NFW profile,
we obtain its weak-lensing-determined $(c_i,M_i)$. From the set of $(c_i,M_i)$ with $i=1, ..., N$, we then derive constraints on the $c$--$M$ relation.

\begin{figure*}
  \centering
  \includegraphics[width=0.75\textwidth]{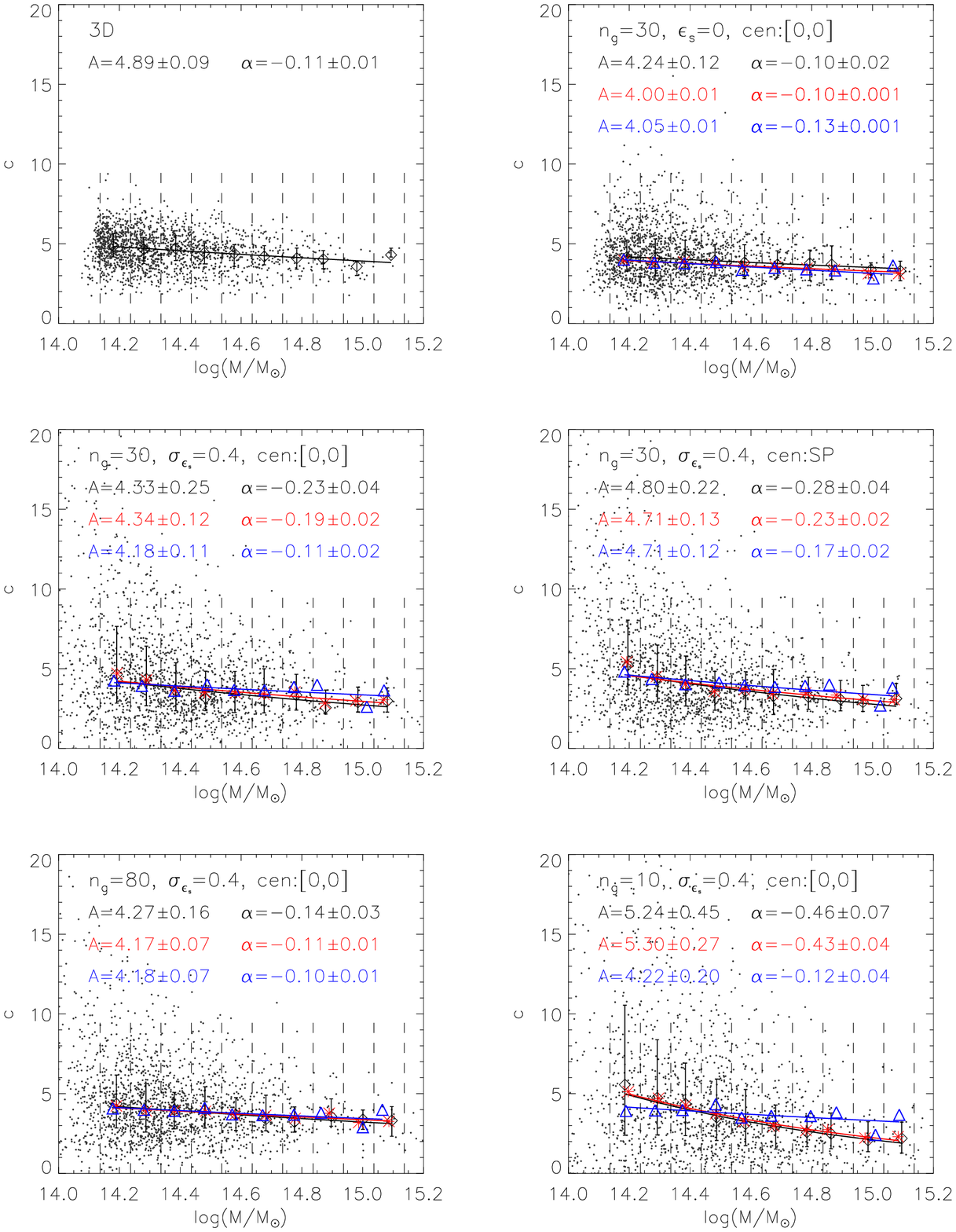}\\
  \caption{ $c$--$M$ relation derived for different cases. The upper-left panel shows the 3D fitting result. The upper-right panel presents the fitting to the 2D reference data with $n_g=30$ without noise. The middle two panels are for the default case using the true center (left) and the SP center (right) for each individual weak-lensing analysis, respectively. The lower-left panel is the fitting results with the noise level of $n_g=80$ and $\sigma_{\epsilon_s}=0.4$ using the true centers. The lower-right panel shows the case with a larger noise level of $n_g=10$ and $\sigma_{\epsilon_s}=0.4$ using the true centers. The small black dots in each panel are the best-fit results of individual weak-lensing analysis in the considered cases. The black lines show the $c$--$M$ relation fitting to the black diamonds, which mark the median concentration and median mass of the black dots in each bin. The red asterisks are from the density profile fitting to the stacked weak-lensing data in each weak-lensing-derived mass bin and red lines present the corresponding $c$--$M$ fitting results. Blue triangles show the density profile fitting results to the stacked weak-lensing data within each true mass ($\mt$)-based bin, and blue lines present the corresponding $c$--$M$ relation fitting results. The error bars are for the range between the first and third quartiles around the median values. The fitted values of $A$ and $\alpha$ are shown in the same colors as the corresponding symbols and lines in each panel.
}\label{fig:cm_relation}
\end{figure*}

Figure \ref{fig:cm_relation} shows the $c$--$M$ plots from our weak-lensing NFW fitting of $1690+66$ clusters for different cases as indicated
in different panels.
In each panel, the black dots represent the best-fit value of $(c, M)$
from individual weak-lensing reduced shear map for each cluster. We divide the mass into $10$ bins from $10^{14}\hh\msun$ to $10^{15}\hh\msun$.
The black diamonds and error bars are the corresponding median values and $[1/4, 3/4]$ percentiles within different mass bins.
The red asterisks and the blue triangles are for the results of $(c, M)$ obtained from NFW fitting to the stacked weak-lensing signals within
each mass bin, where the binning is based on the weak-lensing-derived mass and the true mass of halos $M_{200}$, respectively.

Visually, we see that the apparent $c$--$M$ relation from the binned data (symbols) depends on the binning methods and is also sensitive to the
noise level. For quantitative studies, in the following, we describe two different methods to extract the $c$--$M$ relation from the data.
One is the simple $\chi^2$ fitting often adopted in weak-lensing observational analyses. The other is the Bayesian method taking into
account the scatter and covariance of weak-lensing determined $c$ and $M$.

\subsubsection{Simple $\chi^2$ Fitting Method}

We first adopt the simple $\chi^2$ fitting method that is often employed in weak-lensing observational analyses.
In this approach,  the $c$--$M$ relation is derived by minimizing the following $\chi^2$ function

\begin{equation}\label{eq:cmfit}
\chi^2=\sum_i\frac{({\rm log}\,c_i -{\rm log}\,c_{i,{\rm model}})^2}{\sigma_i^2},
\end{equation}
where $c_i$ is the observed value, $c_{i,{\rm model}}$ is the model prediction assuming the mass is
the weak-lensing-measured mass $M_i$,  and $\sigma_i$ is the error for ${\rm log}\,c_i$.
Specifically, taking the power law relation of Eq.(\ref{eq:cm}), $c_{i,{\rm model}}$ is modeled by $c_{i,{\rm model}}=A(M_i/M_p)^{\alpha}$, where
$M_i$ is set to be the mass determined from weak-lensing measurements, $M_p$ is the chosen pivot mass, and $(A, \alpha)$ are the two free parameters
to be estimated \citep[e.g.,][]{2010PASJ...62..811O,2012MNRAS.420.3213O,2013MNRAS.428.2921D}.

For illustrations, the fitting results to the binned data (symbols) in Figure \ref{fig:cm_relation} are shown by the corresponding lines.
The fitted values of $(A,\alpha)$ are given in each panel. For the data shown by the black diamonds, obtained by directly taking the median values
of the individual measurements (black dots) within different bins (referred to as the direct method),
their corresponding errors are calculated by $\sigma^i_{{\rm log}\,c} /\sqrt{N_i}$, where $\sigma^i_{{\rm log}\,c}$ and
$N_i$ are the $rms$ of ${\rm log}\,c$ and the number of data points within the $i$-th mass bin.
For the red asterisks and blue triangles, $\sigma_i$ is taken to be the measurement error
in ${\rm log}\,c$ from the stacked reduced shear signals in the corresponding $i$-th bin, assuming the error is Gaussian in log-space.

For the 3D case, shown in the upper left panel, $A\sim 4.89$, $\alpha\sim -0.11$, and the scatter $\sigma_{{\rm log}\,c}\sim 0.12$,
fully consistent with the results from \citet{2007MNRAS.381.1450N}.
For the reference 2D case of $(c_{\rm 2D}, M_{\rm 2D})$ without noise (upper right),
the three fitting results for the slope $\alpha$ are similar and in good accord with
the 3D result. For $A$, the 2D results are somewhat lower than the 3D result, in agreement with the
negative bias due to the projection effects shown in Figure \ref{fig:Qmqc_mvir}.
The scatter of $c$ is $\sigma_{{\rm log}\,c}\sim 0.17$, somewhat larger than that of the 3D case.

The results shown in the middle and lower panels in Figure \ref{fig:cm_relation} demonstrate that the derived $c$--$M$ relation
is sensitive to the noise level; the larger the noise, the steeper the slope $\alpha$ if
the weak-lensing-derived mass is used in binning the data (results shown in black and red).
On the other hand, if the binning is based on the true mass of halos (blue),
the best-fit $c$--$M$ relation is nearly independent of the noise level, and the slope $\alpha$ is very consistent with that
from the 2D reference case. The comparison between the left and right middle panels shows that
using different identified centers in the NFW fitting can mildly affect the derived $c$--$M$ relation.

\begin{figure}[!ht]
  \centering
  \includegraphics[width=0.4\textwidth]{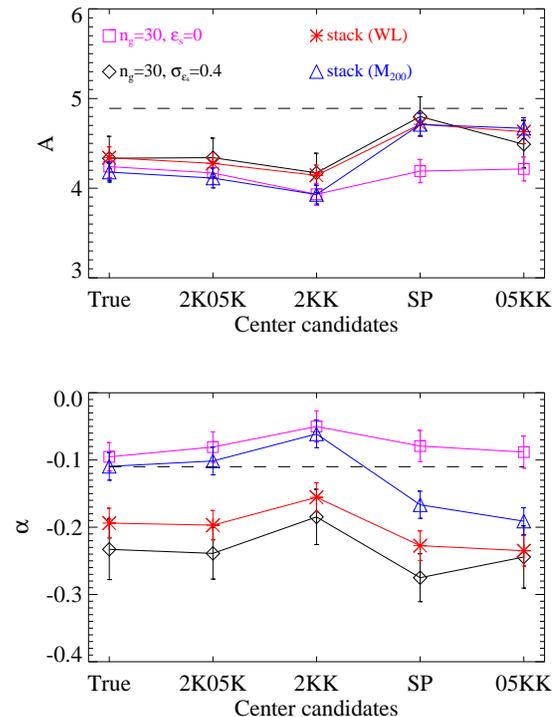}\\
  \caption{Values of derived $A$ and $\alpha$ for the $c$--$M$ relation as a function of different identified centers. The magenta squares are for the 2D results without noise. The blue triangles and red asterisks are for the results from the stacked weak-lensing analyses with the mass binning based on the true mass and the weak-lensing (WL)-derived mass, respectively. Black diamonds are for results of direct fitting. The black, red and blue symbols are all for the default case with $n_g=30\hbox{ arcmin}^{-2}$ and $\sigma_{\epsilon_s}=0.4$. The dashed lines indicate the 3D derived values of $A$ and $\alpha$.}\label{fig:offset_A_ap}
\end{figure}

Figure \ref{fig:offset_A_ap} systematically shows the dependence of the amplitude $A$ (upper) and the slope $\alpha$ (lower) of the $c$--$M$ relation derived from
the $1690+66$ weak-lensing analyzed clusters on the center candidates used in the NFW fitting. Different centers are identified for the default case,
where $n_g=30\hbox{ arcmin}^{-2}$ and $\sigma_{\epsilon_s}=0.4$. The black diamonds, red asterisks, and blue triangles are the corresponding
fitting results from the direct method and stacking methods based on weak-lensing-derived mass and true mass, respectively.
The dashed lines indicate the 3D results. The magenta squares show the results of using the centers identified from the default case but
without including noise in the profile fitting for halos. In this case, we see that the center offset does not significantly affect the derived $c$--$M$ relation.
For 2KK, the value of $A$ is somewhat smaller by $\sim 0.5$, and $|\alpha|$ is smaller by $\sim 0.05$ than that of the case using
the true center in the profile fitting. This is because 2KK results in a large fraction of offset in the range $(0.5r_s,\,r_s)$ for lower mass halos
(see Figure \ref{fig:Peaks_hor}), which leads to a statistically negative bias in the concentration parameter (see Figure~\ref{fig:Q_cm}).

For the cases with noise and data binning based on the weak-lensing-determined mass (black and red),
overall, the slope $\alpha$ is significantly steeper than the results without noise (magenta squares).
There is a mild tendency of larger $|\alpha|$ and $A$ for using centers identified by SP and 05KK than that
using other center candidates.
This is because for SP and 05KK, the identified centers are more sensitive to small-scale structures than the other two methods.
For high-mass halos, the small-scale structures are likely to be real subclumps. On the other hand, however, they correspond
mainly to noise peaks for low-mass halos. Therefore, by including noise in the NFW fitting, low-mass halos are affected more
than high-mass halos compared to the case without noise. This can lead to somewhat higher A and $\alpha$ in the
derived $c$--$M$ relation.
For stacking based on the true mass of halos (blue triangles), the results corresponding to 2K05K and 2KK are consistent with that of
the case without noise (magenta squares). However for SP and 05KK, the effects from small-scale structures caused by noise discussed above
can still be seen here. In summary, comparing to the noise effects, the impact of center offsets is subdominant in the
$c$--$M$ relation studies. The 2K05K method performs the best in center identifications,
and for $E_{\rm SN}>10$, the derived $c$--$M$ relation is about the same as that using the true centers of halos in the NFW fitting.

\begin{figure*}
  \centering
  \includegraphics[width=0.8\textwidth]{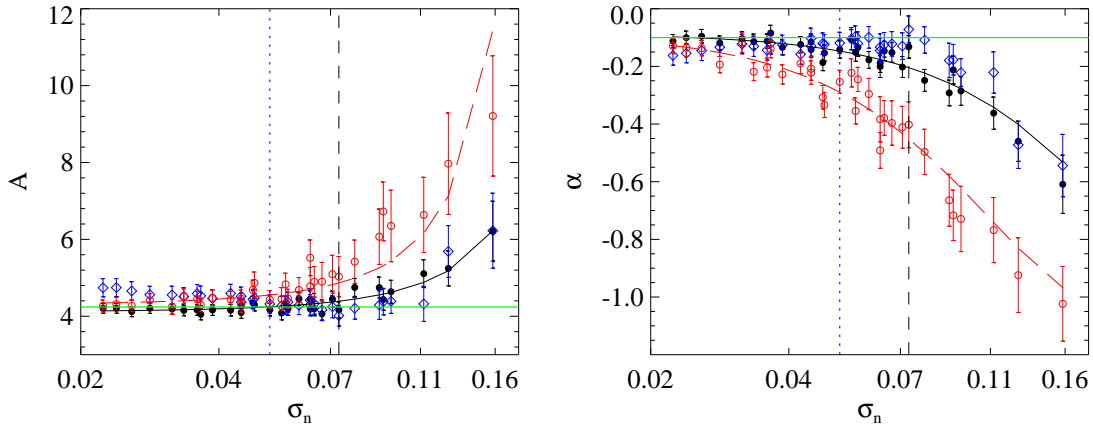}\\
  \caption{Weak-lensing-derived $A$ and $\alpha$ as a function of the noise level $\sigma_n$. The black dots with error bars are the best-fit values from the ``direct method" as shown in Figure \ref{fig:cm_relation}. The blue squares are the fitting results based on the method of \citet{2012MNRAS.421.1073B}. The red circles are the same as black dots but with fitting range from $1\arcmin$ to $10\arcmin$. Black solid lines show the best-fit trend for black dots from Equations (\ref{eq:a_sign}) and (\ref{eq:fap_sign}). The red solid lines are for the red circles. The vertical black dashed and blue dotted lines mark the noise levels of our default case and \citet{2012MNRAS.421.1073B}, respectively. Horizontal green lines are the 2D reference for comparison. The pivot mass in this figure is $M_p/M_{14}=1$ with ${M}_{14}=10^{14}h^{-1}\msun$.}\label{fig:sign_AalpC}
\end{figure*}

We now look at the noise effects on $c$--$M$ relation. The results for 32 noise level catalogs are shown in Figure \ref{fig:sign_AalpC}.
The horizontal axes are for the noise level $\sigma_n=\sigma_{\epsilon_s}/\sqrt{n_g}$.
For each set of analyses with a given $\sigma_n$,
we group the weak-lensing-derived $(c, M)$ according to $M$. Here, we use the weak-lensing-derived mass to bin the data
to resemble weak-lensing cluster analyses in which we do not know the true mass of the clusters.
With the direct method, we fit the median values of $c$ within each mass bin to obtain the amplitude $A$ and
the slope $\alpha$ for the $c$--$M$ relation. The results are shown by the black dots with error bars in Figure \ref{fig:sign_AalpC}.
The green horizontal line in each panel indicates the corresponding 2D reference result. It is seen clearly that both the derived $A$ and $\alpha$ monotonically
depend on the noise level $\sigma_n$, the larger the noise, the higher the amplitude $A$ and the steeper the slope $\alpha$.

The dependence of $\alpha$ on $\sigma_n$ can be well described by the following functional form
\begin{equation}\label{eq:fap_sign}
\alpha(\sigma_n)=e^{-\frac{\sigma_n^2}{2\sigma_{\alpha}^2}}-\alpha_0,
\end{equation}
where $(\alpha_0,\,\sigma_{\alpha})=(1.09\pm0.01,\,0.15\pm0.01)$. The fitting result is shown by the black solid line in the right panel of Figure \ref{fig:sign_AalpC}.
Due to its dependence on the pivot mass, $M_p$, the trend of $A$ is different for different choices of $M_p$.
Here, we use $M_p=10^{14}h^{-1}M_{\odot}$, and have
\begin{equation}\label{eq:a_sign}
A(\sigma_n)=e^{\frac{\sigma_n^2}{2\sigma_{A}^2}}+A_0,
\end{equation}
where $(A_0,\,\sigma_{A})=(3.12\pm0.04,\,0.10\pm0.01)$.

The blue squares with error bars in Figure \ref{fig:sign_AalpC} are for the results using the method of \citet{2012MNRAS.421.1073B}
to determine $(c,M)$ for each individual map, in which the reduced tangential shears of all the source galaxies
(rather than the radially binned signals) in the radial range of $[0\overset{\prime}{.}5, 15\arcmin]$ (rather $\ge 1\arcmin$)
around the center are used to perform the NFW fitting. We see that for $A$, the results have a trend similar to those of black dots.
For $\alpha$, the blue ones are somewhat flatter than the black ones for noise levels larger than 0.04
but the dependence on $\sigma_n$ is still clearly seen. Considering the noise level used in \citet{2012MNRAS.421.1073B}
with $\sigma_n=0.2\sqrt{2}/\sqrt{30}\approx 0.052$ (dotted vertical line), the result shown by the blue symbol here is consistent with their result.
We also show the $A-\sigma_n$ and $\alpha-\sigma_n$ relation (red circles with error bars in Figure \ref{fig:sign_AalpC})
for the result with fitting range from $1\arcmin$ to $10\arcmin$. The red solid lines are the fitting results using Equations (\ref{eq:fap_sign}) and (\ref{eq:a_sign}).
It is seen that the noise effect is larger for this smaller fitting range.

The results presented in Figure \ref{fig:sign_AalpC} show that, given the same set of clusters, the simple fitting approach adopted here can lead to a
different $c$--$M$ relation depending sensitively on the noise level.
This indicates that such a fitting method is inadequate, and a more sophisticated method, taking into account the scatter
of both $c$ and $M$, and particularly their correlations, is needed in order to extract the unbiased $c$--$M$ relation.

\subsubsection{Bayesian Method}

Technically, the bias effect of noise on the $c$--$M$ relation shown above arises because
the weak-lensing-derived mass is used directly as the true mass to calculate the model predicted $c_{i, model}$ in Equation~(\ref{eq:cmfit}).
On the other hand, however, errors exist in weak-lensing-derived mass. More importantly, errors in mass are strongly
correlated with errors in the concentration parameter.

\begin{figure*}[tp]
  \centering
  \includegraphics[width=0.8\textwidth]{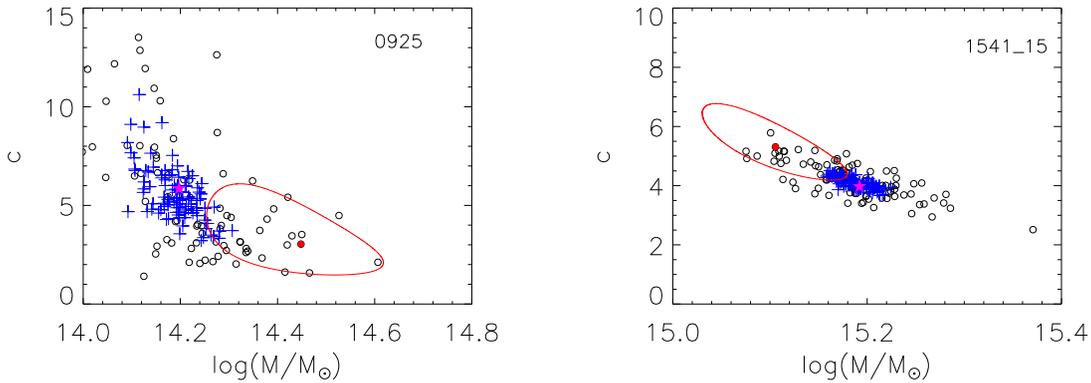}\\
  \caption{Scatter in weak-lensing determined $(c, M)$ for two clusters, one from a low-mass sample (left) and one from a high-mass sample (right).
Two noise levels are considered with $n_g=30 \hbox{ arcmin}^{-2}$ and $\sigma_{\epsilon_s}=0.4$ (black circles), and $n_g=80\hbox{ arcmin}^{-2}$
and $\sigma_{\epsilon_s}=0.2$ (blue plusses), respectively. At each noise level, we generate $100$ mock weak-lensing observations
for each of the two clusters with different realizations of the intrinsic ellipticities of background galaxies.
The black circles/blue plusses are the best-fit results of $(c, M)$ from individual realizations.
The magenta symbol in each panel indicates the reference $(c_{\rm 2D}, M_{\rm 2D})$ without noise.
The red symbol in each panel shows the result from one particular realization and the red contour is the corresponding
$1\sigma$ measurement error range.}\label{fig:cm1556}
\end{figure*}

In Figure \ref{fig:cm1556} we show two examples, one for a low-mass halo (left) and one for a high-mass halo (right).
For each halo, two noise levels of $\sigma_n$ are considered. For each $\sigma_n$, we generate $100$ weak-lensing data sets for
each map by making $100$ realizations for the background galaxy distribution and intrinsic ellipticity assignments.
The best-fitted $(c, M)$ from each of the $100$ mock weak-lensing analyses is shown in Figure \ref{fig:cm1556},
where the black circles and blue plusses are for high and low noise levels, respectively.
The reference value of $(c_{\rm 2D}, M_{\rm 2D})$ is shown as the magenta five-pointed star.
The strong correlation between the scatter in $c$ and in $M$ from different realizations is clearly seen.
For a high noise level, the scatters extend to a large area along the degenerate direction.
Considering a specific ``observation'', the derived mass (red filled dot) can be higher than $M_{\rm 2D}$ (magenta star), as
shown in the left panel, and the corresponding $c$ is then lower than $c_{\rm 2D}$. For this measurement, the $1\sigma$
confidence region around the best-fit point is shown by the red contour around the red dot. Another measurement giving rise to
a mass that is lower than the reference value and a concentration higher than the reference one is
shown in the right panel for the high-mass halo. Then, when we group $(c,M)$ based on the weak-lensing-derived mass, a halo with a low/high true mass
can be grouped into a high/low weak-lensing-derived mass bin. For those halos, their derived concentrations from NFW fitting are
systematically lower/higher than their underlying true concentrations. Furthermore, because the degeneracy between $c$ and $M$
shown in Figure \ref{fig:cm1556} is significantly steeper than the true $c$--$M$ relation for dark matter halos,
thus the resulting $c$--$M$ relation derived by the simple $\chi^2$ fitting method is affected considerably by this degeneracy and
can be much steeper than the true relation.

Here, we apply a Bayesian method properly taking into account the scatter in $c$ and $M$ and their covariance.

In this method, we model the probability distribution of the observed $(c_{\rm ob},M_{\rm ob})$ given the true $(c_T,M_T)$ at the noise level
$\sigma_n$ as the 2D Gaussian distribution in log-space. It is noted that the true $(c_T,M_T)$ here corresponds to the
reference weak-lensing 2D analyses $(c_{\rm 2D}, M_{\rm 2D})$ without noise. The distribution is written as

\begin{equation}\label{eq:pcm_onT}
P(c_{\rm ob},M_{\rm ob}|c_\smT,M_\smT,\sn)=\frac{\displaystyle {\rm exp}(-\frac{1}{2}\mathcal{X}
\mathcal{C}^{-1}\mathcal{X}^{T})}{\displaystyle 2\pi\sigma_\smM\sigma_c\sqrt{1-r_{c\smM}^2}},
\end{equation}
where
\begin{displaymath}
\begin{array}{ll}
\mathcal{X}={\displaystyle(\lgg\frac{M_{\rm ob}}{M_\smT}\quad\lgg\frac{c_{\rm ob}}{c_\smT})}, \nonumber\\\\
\mathcal{C}=
\left(
              \begin{array}{cc}
                \sigma_{\smM}^2 & r_{c\smM}\sigma_{\smM}\sigma_{c} \\
                r_{c\smM}\sigma_{\smM}\sigma_{c} & \sigma_{c}^2  \\
              \end{array}
            \right).\\
\end{array}
\end{displaymath}
Here, $\sigma_\smM$ and $\sigma_c$ are the standard deviations of $\lgg M$ and $\lgg c$ with respect to the true values, respectively;
$r_{c\smM}$ is the Pearson's correlation coefficient between $\lgg M$ and $\lgg c$;
$\mathcal{X}^{T}$ denotes the transposed vector $\mathcal{X}$; and $\mathcal{C}^{-1}$ is the inverse of the covariance matrix, $\mathcal{C}$.

As seen in the upper right panel of Figure \ref{fig:cm_relation}, given an $M_T$, scatter exists in the value of $c_T$.
Taking into account the scatter, we can write
\begin{equation}\label{eq:pcm_onMT}
\left.
 \begin{array}{l}
P(c_{\rm ob},M_{\rm ob}|M_\smT,\sn)=\\\\
\int P(c_{\rm ob},M_{\rm ob}|c_\smT,M_\smT,\sn)P(c_\smT|M_\smT){\it d}c_\smT, 
\end{array}\right.
\end{equation}
and $P(c_\smT|M_\smT)$ is modeled approximately by
\begin{equation}
P(c_\smT|M_\smT){\it d}c_\smT=\displaystyle\frac{1}{\sqrt{2\pi}\sigma_{\rm in}}{\rm exp}\left[-\frac{\lgg^2\frac{c_\smT}
{\langle c_\smT\rangle}}{2\sigma_{\rm in}^2}\right]{\it d}\lgg\,c_\smT,
\end{equation}
where $\langle c_\smT\rangle$ and $\sigma_{\rm in}$ are the median value and the intrinsic dispersion of $c_T$ given $M_T$, respectively.
For $\langle c_\smT\rangle$, it is determined directly from $M_T$ by using the $c$--$M$ relation.
It is this relation that we aim to extract from a sample of weak-lensing studied clusters.

From above, the probability of $(c_{\rm ob},M_{\rm ob})$ given a $\sn$ can be written as

\begin{equation}\label{eq:pcm}
P(c_{\rm ob},M_{\rm ob}|\sn)=\frac{\int_{{M}_{\rm low}}^{\infty} P(c_{\rm ob},M_{\rm ob}|M_\smT,\sn){\it d}{\rm n}(M_\smT)}{\int_{{M}_{\rm low}}^{\infty}{\it d}{\rm n}(M_\smT)}, 
\end{equation}\\
and further the probability of $c_{\rm ob}$ given $(M_{\rm ob},\sn)$ as
\begin{equation}\label{eq:pc_onM}
P(c_{\rm ob}|M_{\rm ob},\sn)=\frac{\int_{{M}_{\rm low}}^{\infty} P(c_{\rm ob},M_{\rm ob}|M_\smT,\sn){\it d}{\rm n}(M_\smT)}{\int_{{M}_{\rm low}}^{\infty} P(M_{\rm ob}|M_\smT,\sn)
{\it d}{\rm n}(M_\smT)}, 
\end{equation}\\
where $M_{\rm low}$ is the lower mass limit for integration showing the mass-limited selection effect, and ${\rm n}(M_\smT)$ is the halo mass function
taken from \citet{2008ApJ...688..709T} (see the Appendix).
The probability of $M_{\rm ob}$ given $(M_\smT,\sn)$ is
\begin{equation}
P(M_{\rm ob}|M_\smT,\sn)=\int P(c_{\rm ob},M_{\rm ob}|M_\smT,\sn){\it d}c_{\rm ob}.
\end{equation}

From Equation (\ref{eq:pc_onM}), we then obtain the theoretically expected median concentration $\langle c_{\rm ob,model}\rangle$ given a
$c$--$M$ relation for $\langle c_T\rangle$ and $M_T$ by
\begin{equation}\label{eq:md_c}
\int_{\langle c_{\rm ob,model}\rangle}^{\infty}P(c_{\rm ob}|M_{\rm ob},\sn){\rm d}c_{\rm ob}=\frac{1}{2}.
\end{equation}
With a large observational sample of weak-lensing studied clusters, we can fit $\langle c_{\rm ob,model}\rangle$ to the median value of
the observed $c$ of different $M_{\rm ob}$ bins to extract the underlying $c$--$M$ relation. It is done by minimizing the $\chi^2$ defined by
\begin{equation}\label{eq:bcmfit}
\chi^2=\sum_i\frac{({\rm log}\,c_i -{\rm log}\,\langle c_{\rm ob,model}\rangle_{i})^2}{\sigma_i^2}.
\end{equation}
Comparing to Equation (\ref{eq:cmfit}), $\langle c_{\rm ob,model}\rangle_{i}$ is calculated by Equation (\ref{eq:md_c}), rather than by the simple $c$--$M$ relation with
$M_{\rm ob}$ being assumed to be the true mass.

For small samples with a limited number of clusters, we can derive the $c$--$M$ relation parameters $(A,\alpha)$ by maximizing the likelihood
\begin{equation}\label{eq:maxL}
{\rm ln}\mathcal{L}=\displaystyle \sum_i {\rm ln}\,P_i(c_{\rm ob},M_{\rm ob}|\sn)
\end{equation}
where $i$ marks the $i$-th observed cluster.

In the Bayesian approach here, the scatter and covariance of $c_{\rm ob}$ and $M_{\rm ob}$ given $M_T$ are taken into account in the matrix $\mathcal{C}$.
We calculate $\mathcal{C}$ with our simulated mock weak-lensing analyses for all of the $1756$ clusters by considering $18$ different
noise levels with $\sn$ in the range of $[0.03, 0.2]$.
For each cluster at each noise level, $100$ mock observations are generated, similar to that shown in Figure \ref{fig:cm1556}.
Based on the $100$ best-fit values of $(c_{\rm ob},M_{\rm ob})$ for each $(c_\smT,M_\smT)$, the corresponding scatter $(\sigma_\smM,\sigma_c)$
and the correlation coefficient $r_{r\smM}$ are calculated. From the left panel of Figure \ref{fig:Qmqc_mvir}, it is seen that
$M_{\rm 2D}$ (i.e., $M_\smT$ here) is very close to the true mass of halos $\mt$. We thus use $M_\smT=M_{200}$ in the calculations here.
For $c_\smT$, we set $c_\smT=c_{\rm 2D}$, keeping in mind that the 2D reference value $c_{\rm 2D}$ is systematically lower than that of the 3D halos,
$c_{\rm 3D}$, by about $10\%$.
We then calculate the median values of $\sigma_\smM$, $\sigma_c$ and $r_{c\smM}$ within different mass bins. Their quantitative dependencies on
the mass $M_\smT$ and the noise level $\sigma_n$ are presented in the Appendix.

In Figure \ref{fig:covMT}, we show examples of their dependencies on $\sigma_n$ (top panels) and mass (bottom panels).
The black and red lines are the NFW fitting range from $1\arcmin$ to $15\arcmin$ and from $1\arcmin$ to $10\arcmin$, respectively.
It is seen clearly that $\sigma_\smM$, $\sigma_c$, and correlation coefficient $r_{c\smM}$ are increasing functions of the noise level $\sigma_n$,
and decreasing functions of mass. This indicates, in agreement with the results presented in Section 4.2.1, that
the derived $c$--$M$ relation from a sample of weak-lensing studied clusters suffers from the noise effects if they are not properly taken into consideration.
The dependencies of $(\sigma_\smM, \sigma_c, r_{c\smM})$ on the mass indicate the effects of sample selections.
Using a narrower range  in the NFW fitting (red lines), both $\sigma_\smM$ and $\sigma_c$ are larger, and $r_{c\smM}$ is more negative.
This explains the larger bias in the $c$--$M$ relation from the simple $\chi^2$ fitting shown in Figure \ref{fig:sign_AalpC} (red symbols).

\begin{figure*}[tp]
  \centering
  \includegraphics[width=0.8\textwidth]{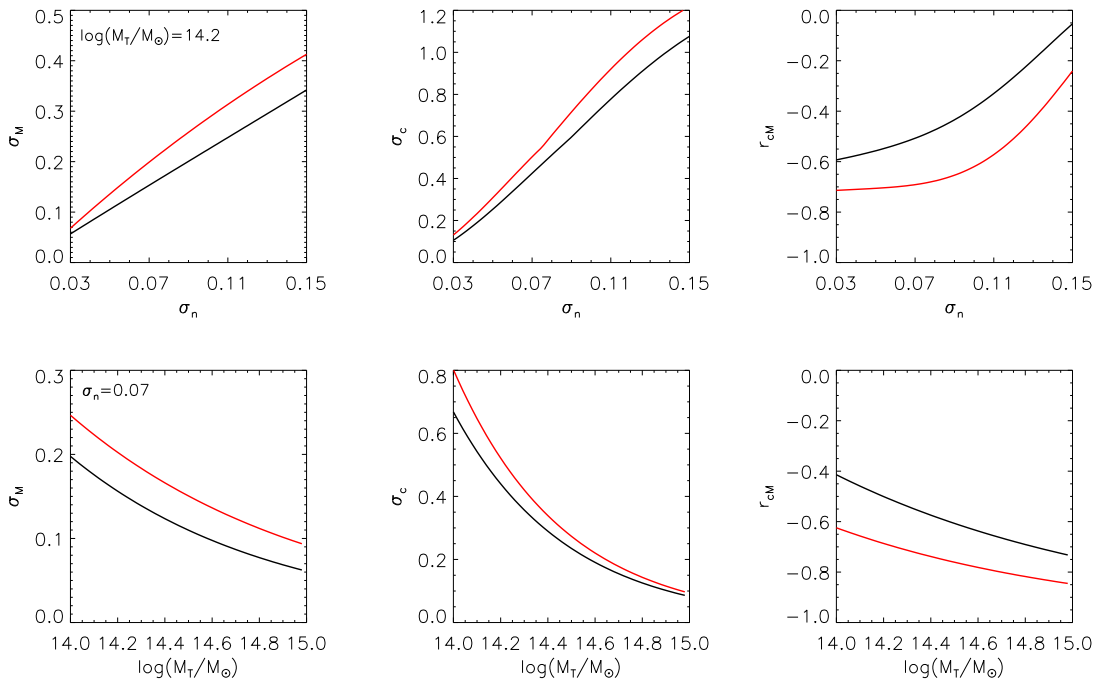}\\
  \caption{Dependence of the scatter $(\sigma_\smM,\sigma_c)$ and correlation coefficient $r_{c\smM}$ on the noise level $\sigma_n$ (top panels)
and mass (bottom panels). In the top panels, a halo with mass $M_\smT=10^{14.2}\msun$ is considered.
In the bottom panels, $\sigma_n=0.07$, close to our default case.
The black lines are for the NFW fitting range from $1\arcmin$ to $15\arcmin$, and the range for the red lines is $1\arcmin$ to $10\arcmin$ .}\label{fig:covMT}
\end{figure*}

\begin{figure*}[tp]
  \centering
  \includegraphics[width=0.8\textwidth]{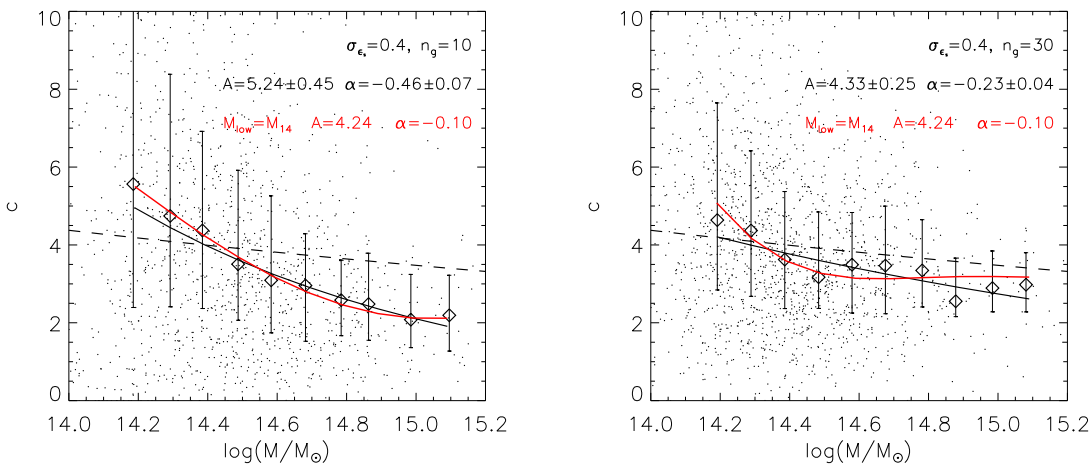}\\
  \caption{Apparent $c$--$M$ relation. The left is for the noise level with $\sigma_{\epsilon_s}=0.4$ and $n_g=10$.
The right panel corresponds to $\sigma_{\epsilon_s}=0.4$ and $n_g=30$. The small black dots in each panel are the best-fit results from weak-lensing analysis for
individual clusters. Black diamonds denote the median concentration and median mass of the black dots in each mass bin.
The error bars are for the range between the first and third quartiles around the median values.
Black solid lines are the fitting results from the simple $\chi^2$ fitting method with the best-fit values of
$A$ and $\alpha$ listed in black. The red lines are predicted from our Bayesian method by
fixing $A=4.24$, $\alpha=-0.10$, and $M_{\rm low}=M_{14}$. The dashed line in each panel represents the $c$--$M$ relation of the 2D reference case
with $A=4.24$ and $\alpha=-0.1$.}
\label{fig:BayP}.
\end{figure*}

To demonstrate our Bayesian approach, in Figure \ref{fig:BayP}, we show the apparent median $c$--$M$ relation
(\ie $(\langle c_{\rm ob} \rangle-M_{\rm ob})$ relation) at different noise levels, expected theoretically from our Bayesian analyses, by using
the underlying $c$--$M$ relation with $A=4.24$, and $\alpha=-0.10$ in relating $\langle c_T\rangle$ to $M_T$ (red lines).
The lower mass limit is taken to be $M_{\rm low}=1\times10^{14}h^{-1}\msun$. The black dots are the fitted $(c,M)$ from individual clusters, and the
black diamonds are the corresponding median values within different weak-lensing-determined mass bins. The black solid line is
for the $c$--$M$ relation obtained by using the simple $\chi^2$ fitting method to the median values (diamonds)
with the corresponding best fit values of $A$ and $\alpha$ listed in black. The black dashed line in each panel
shows the $c$--$M$ relation obtained in the 2D weak-lensing reference case without noise, where $A=4.24$, and $\alpha=-0.10$.
It is seen clearly that our Bayesian predictions agree with the data (black diamonds) very well; the larger the noise, the steeper
the apparent $\langle c_{\rm ob} \rangle-M_{\rm ob}$ relation. Put another way, the results here show that by using the Bayesian method,
we can expectedly extract the underlying $c$--$M$ relation unbiasedly from the noise affected data.

\begin{figure*}
  \centering
  \includegraphics[width=0.8\textwidth]{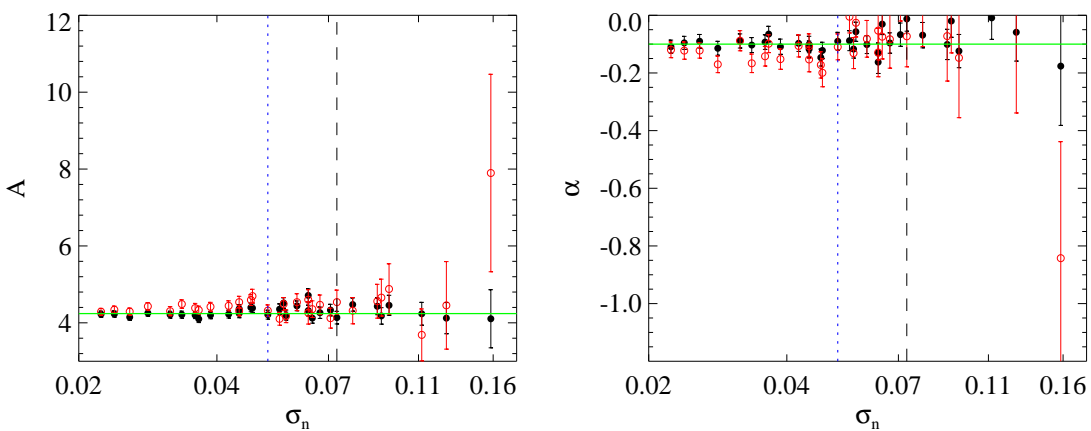}\\
  \caption{Weak-lensing-derived $A$ and $\alpha$ as a function of the noise level $\sigma_n$ from our Bayesian method.
The black and red symbols are for fitting ranges $(1\arcmin, 15\arcmin)$ and $(1\arcmin, 10\arcmin)$, respectively.}\label{fig:sign_AalpB}
\end{figure*}

With our Bayesian method, we refit the 32 noise level large samples (the same samples used in deriving the results of Figure \ref{fig:sign_AalpC})
using Equation (\ref{eq:bcmfit}).
In the fitting, $M_{\rm low}$ is fixed to be $1\times 10^{14}h^{-1}\msun$, consistent with our simulation samples.
The results for the derived $A$ and $\alpha$ are presented in Figure \ref{fig:sign_AalpB}. The colors of the symbols have the same meanings as that
in Figure \ref{fig:sign_AalpC}. We then see very clearly that with the Bayesian method, we can indeed extract the unbiased $c$--$M$ relation
(with respect to the 2D reference case) successfully for a wide range of noise levels. This demonstrates the great potential
to probe the underlying $c$--$M$ relation of dark matter halos using large samples of weak-lensing studied clusters, keeping in mind the
mild difference in $A$ between the 2D reference case and that of the 3D halos. It is noted that when $\sigma_n$ is too large,
the Gaussian approximation used in Equation (\ref{eq:pcm_onT}) may not be valid, and that could introduce a certain level of bias in the derived $c$--$M$ relation.

\subsection{Monte Carlo Analyses for Samples with a Limited Number of Clusters}

The results shown in Figures \ref{fig:sign_AalpC} and \ref{fig:sign_AalpB} are obtained by analyzing $1756$ simulated clusters.
On the other hand, current observational weak-lensing cluster analyses are still limited to few tens of clusters. For these,
the sample variance can be significant. Here, we perform Monte Carlo studies for samples similar to Ok10 and Og12.

First, we note that the simple $\chi^2$ fitting method, \ie Equation(\ref{eq:cmfit}), is adopted by both Ok10 and by Og12 when deriving the $c$--$M$ relation.
The two consider $\sigma_i$ somewhat differently. Ok10 include only the measurement errors with $\sigma_i=\sigma_{\rm me}$, but Og12 also take into account
the intrinsic dispersion $\sigma_{\rm in}$. Here for consistent comparisons, we follow Og12 by using $\sigma_i^2=\sigma_{\rm me}^2+\sigma_{\rm in}^2$ for both Ok10 and Og12
Monte Carlo analyses. We take $\sigma_{\rm in}=0.12$ (Og12).
We also convert their given $(c, M)$ corresponding to the overdensity $\Delta_{\rm vir}(z)=18\pi^2+82[\Omega_m(z)-1]-39[\Omega_m(z)-1]^2$ \citep{1998ApJ...495...80B}
to $(c, M)$ corresponding to $\Delta=200$ for the NFW profile \citep[e.g.,][]{2003ApJ...584..702H,2012MNRAS.424.1244F}.

For Ok10, their sample contains a total of $30$ X-ray selected clusters with redshift $z\sim 0.2$.
With high-quality Subaru/Suprime-Cam data, they perform detailed weak-lensing analyses for each of the clusters. To study the $c$--$M$ relation,
they pick out $19$ clusters whose weak-lensing data can be better fitted by the NFW profile than by the singular isothermal sphere (SIS).
With this subsample of clusters, they obtain $A=5.75^{+2.47}_{-1.90}$ and $\alpha=-0.37^{+0.21}_{-0.2}$ using the overdensity
parameter, $\Delta=200$, and the pivot mass, ${M}_p={M}_{14}=1\times 10^{14}h^{-1}\msun$.
For their observations, the number density of source galaxies is in the range of $n_g\sim 10$--$30\hbox{ arcmin}^{-2}$ and their intrinsic ellipticity
parameter $\sigma_{\epsilon_s}\sim 0.4$.

To check the fitting method with the data given in their Table 6, taking care of the conversion of $(c, M)$ corresponding to
the overdensity $\Delta_{\rm vir}$ to that with respect to $\Delta_{200}$, we refit the $c$--$M$ relation using Equation(\ref{eq:cmfit})
with $\sigma_i$ including only the measurement error $\sigma_{\rm me}$ and ${M}_p={M}_{14}$. We find
$A=5.77\pm2.19$ and $\alpha=-0.37\pm0.20$, almost exactly the same as their results for $\Delta=200$. By including $\sigma_{\rm in}= 0.12$ in the fitting, the Ok10 data give rise to $A=7.81\pm3.55$ and $\alpha=-0.44\pm0.25$.
In the following analyses for Ok10, we first consider the simple $\chi^2$ fitting method, including $\sigma_{\rm in}$ in deriving the $c$--$M$ relation.
We then also perform Bayesian analyses using their data and the Monte Carlo subsamples.

For Monte Carlo analyses corresponding to the simple $\chi^2$ fitting method, we generate $1000$ sets of subsamples from our full simulation sample of clusters.
Each set contains $19$ clusters with weak-lensing-derived mass $M\ge2\times10^{14}h^{-1}\msun$ and the reduced $\chi^2\le2$ for the NFW fitting.
To populate source galaxies for a cluster in a subsample, we choose $n_g$ randomly in the range of $[10, 30]\hbox{ arcmin}^{-2}$,
and take $\sigma_{\epsilon_s}= 0.4$. We then fit the 1D reduced tangential shear signals for each cluster to obtain
the weak-lensing determined $(c, M)$.  The default fitting range with $[1\arcmin,15\arcmin]$ is adopted, which is consistent
with the fitting range adopted by Ok10.

\begin{figure*}
  \centering
  \includegraphics[width=0.81\textwidth]{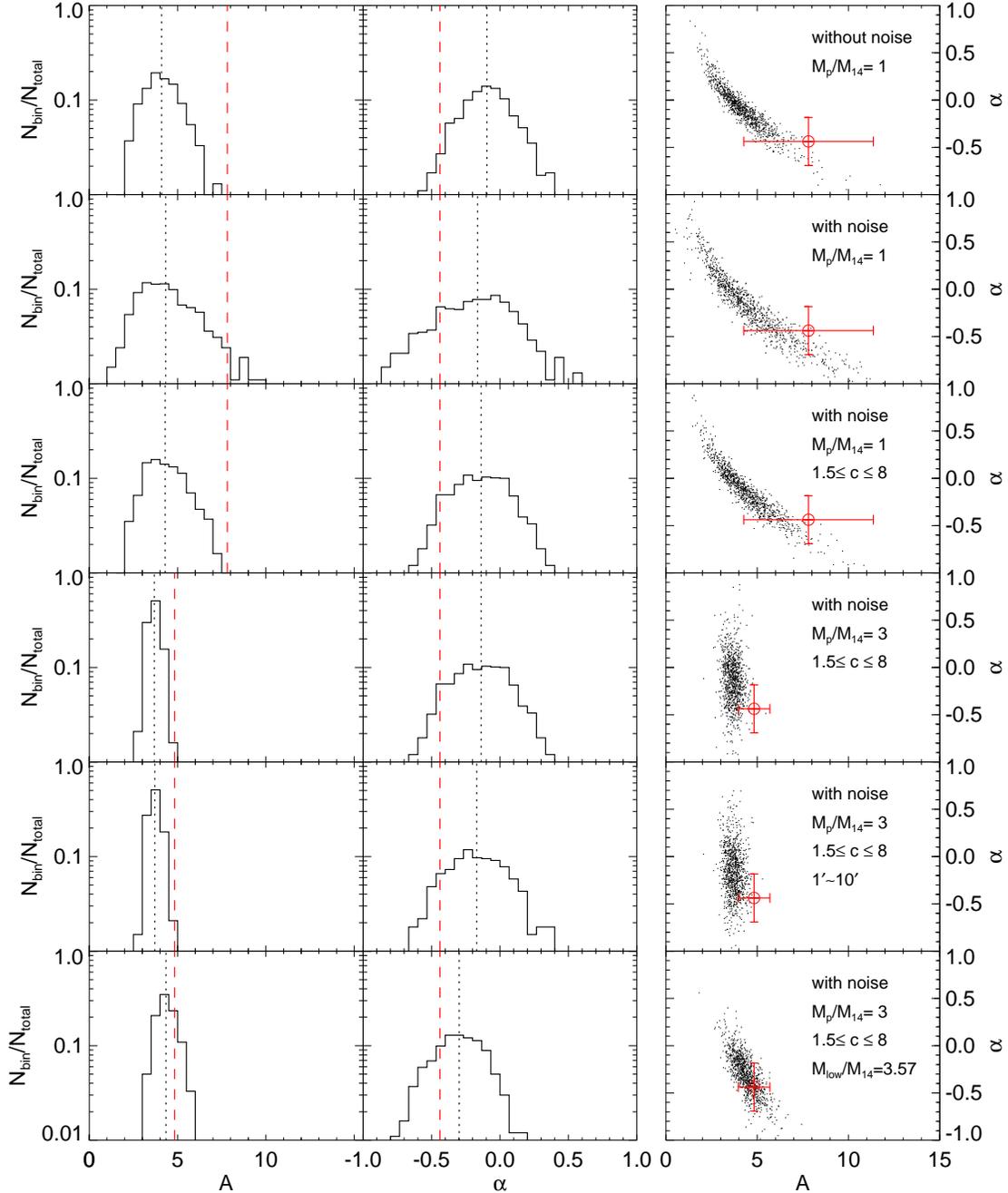}\\
\caption{Results from $1000$ mock realizations compared with the result of Ok10.
The first and second rows are for the cases without and with noise, respectively.
The third row shows the results for clusters with a weak-lensing-derived concentration of $1.5\leq c \leq 8$.
For the fourth row, the data for fitting is the same as that for the third row but the pivot mass
is taken to be $M_p=3\times10^{14}h^{-1}\msun$. The fifth row is for the results using the NFW fitting range of $[1\arcmin,10\arcmin]$.
For the bottom row, a selection condition with $M\ge M_{\rm low}=3.57M_{14}$ is applied to the subsamples.
The left two columns present the probability distributions for $A$ and $\alpha$, respectively.
The vertical dotted lines show the corresponding median values, and the red dashed vertical lines are for the result of Ok10.
The right column shows the results on the $A-\alpha$ plane with the Ok10 result represented
by the red symbols with error bars.}\label{fig:eobs19}
\end{figure*}

The probability distributions for the derived $A$ (left column) and $\alpha$ (middle column) from the $1000$ subsamples are shown
in Figure \ref{fig:eobs19}, where the red dashed vertical lines indicate the best-fit results of Ok10 (including $\sigma_{\rm in}$ in the fitting),
and the black dashed vertical lines
show the median values of the distributions. The right column shows the results on the $A$--$\alpha$ plane with the red circles
with error bars for the result of Ok10. The first and second rows are for the cases without and with noise, respectively. It is seen
that the existence of noise indeed increases $|\alpha|$ and $A$ and make the result of Ok10 agree with the Monte Carlo analyses
better than the case without noise, although the trend is not very large given the considered noise levels.
We notice that, in the sample of Ok10, their weak-lensing-derived $c$ defined by $\Delta_{\rm vir}$ is in the range of $[2, 10]$,
which corresponds approximately to $c\sim [1.5, 8]$ defined by $\Delta=200$. In the third row of Figure \ref{fig:eobs19},
we show the results by adding the constraint of $c=[1.5, 8]$ in generating subsamples. We see that this additional
constraint tends to lower the values of $|\alpha|$ somewhat as expected. The agreement in $\alpha$ between Ok10 and
the Monte Carlo results is still reasonable, but $A$ from Ok10 lies at the tail of the distribution.

For the top three rows, the pivot mass is $M_p=10^{14}h^{-1}\msun$, the same as that used in Ok10.
With this choice, $A$ and $\alpha$ are strongly correlated as seen
from the last columns, indicating that the errors in $A$ and $\alpha$ of Ok10 are not independent
of each other. In the fourth row, we show the results by choosing $M_p=3\times 10^{14}h^{-1}\msun$. In this case,
the correlation between $A$ and $\alpha$ is largely removed, and thus the errors of $A$ and $\alpha$ are nearly
independent, making the comparison of the Ok10 result with the Monte Carlo analyses easier.
From the right panel of the fourth row, we see from taking into account the noise effect, that the slope $\alpha$ in the $c$--$M$ relation obtained
from Ok10 is not significantly different from the result from simulations. On the other hand, the amplitude $A$ is
larger than the simulation result. In the fifth row, we show the result based on the fitting range of $[1\arcmin,10\arcmin]$;
the slope is slightly steeper than the result using the fitting range of $[1\arcmin,15\arcmin]$.

To see the sample selection effect, in the bottom panels of Figure \ref{fig:eobs19}, we show the Monte Carlo results for
subsamples generated from the parent halo sample with true mass $M_{200}\ge3.57M_{14}$.
It is seen that in this case, the agreement between the Ok10's observational result and that of the Monte Carlo simulations improves
significantly. This shows the importance of understanding the sample selection function in making comparisons between observational
and theoretical results. It is interesting to note that the true mass limit $M_{200}\ge3.57M_{14}$ adopted here is obtained by performing
our Bayesian analyses to Ok10's data to be described in the following.

For Bayesian analyses using Ok10's data, we regenerate $10,000$ realizations of the data set.
For each realization, we randomly pick up a point in the $(c,M)$ plane according to the 2D error distribution around
$(c_{\rm ob}, M_{\rm ob})$ given by Ok10 for each cluster.
The error distribution is assumed to be 2D Gaussian in log-space with the dispersions in $c$ and $M$ taken to be their
measurement errors from Ok10. The cross-correlation coefficient for the error distribution is taken from the fitting relation of $r_{c\smM}-M_\smT$
obtained from our simulation analyses with NFW fitting range $[1\arcmin,15\arcmin]$ shown in the Appendix,
where best-fit $M_{\rm ob}$ from Ok10 is used as $M_\smT$.
We then perform a Bayesian fitting using Equation (\ref{eq:maxL}) for every one of the $10,000$ realizations, each with $19$ data points.
The pivot mass $M_p=3M_{14}$ is adopted here.
Note that with $M_p=3M_{14}$, we have $A\simeq3.80$ for our 2D reference case using the MS cluster sample.

As a test, we first fix the underlying $c$--$M$ relation with $(A,\alpha)=(3.80,-0.1)$ and vary only the selection function $M_{\rm low}$ in the fitting.
For each realization of the data set, the best-fit value of $M_{\rm low}$ is recorded. We find that the median value of $M_{\rm low}/M_{14}$
is $3.57_{-0.20}^{+0.21}$ where the errors present the $68\%$ confidence range. This is the
mass limit used in the analyses shown in the bottom panels of Figure \ref{fig:eobs19}.
The likelihood at this median $M_{\rm low}$ is ${\rm ln}\mathcal{L}\sim11.08$ for Ok10's best-fit $(c_{\rm ob}, M_{\rm ob})$.
This indicates that if Ok10 clusters are indeed in accord with the selection function with
$M\ge M_{\rm low}=3.57M_{14}$, the apparent steep $c$--$M$ relation seen in Ok10 is actually largely consistent with the
underlying true $c$--$M$ relation, once the noise effects are properly taken into account in the Bayesian analyses.

\begin{figure*}
  \centering
  \includegraphics[width=0.81\textwidth]{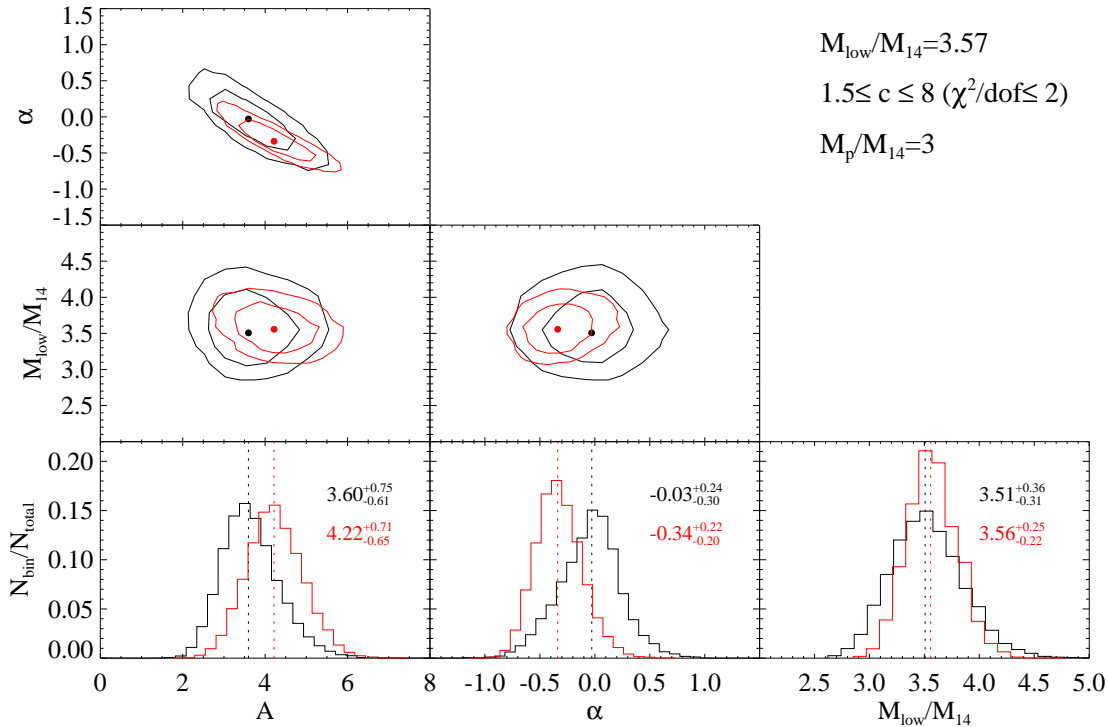}\\
\caption{Probability density contours and distributions from our Bayesian fitting to the 10,000 realizations of the Ok10 data set (red)
and to that of the 10,000 subsamples from the MS sample (black). The contours present isodensity levels enclosing 68\% and 95\% of the 2D
probabilities. The histograms show the 1D probability distribution functions. The dots and dashed lines correspond to the marginal median values.
The criteria for constructing Monte Carlo subsamples from our MS clusters are shown in the top-right corner.}\label{fig:Simok19}
\end{figure*}

We then perform Bayesian fitting with all three parameters $(A, \alpha, M_{\rm low})$ set to be free,
and record their best-fit values from each of the $10,000$ realizations.
After marginalization, we find that the 1D marginal median values are
$A=4.22_{-0.65}^{+0.71}$, $\alpha=-0.34_{-0.20}^{+0.22}$, and $M_{\rm low}=3.56_{-0.21}^{+0.25}$, respectively.
Here, the errors correspond to the $68\%$ confidence ranges.
The likelihood for this set of median values is ${\rm ln}\mathcal{L}\sim11.51$.
In Figure \ref{fig:Simok19}, we show, in red lines, the 2D probability contours (upper rows) and 1D distributions (bottom row)
constructed from the $10,000$ best-fit results.
The red dots in the top two panels and the red lines in the lower panels indicate the corresponding marginal medians.
The contours present 2D isodensity levels enclosing $68\%$ and $95\%$ of the probability of the 10,000 fittings. Histograms in the bottom rows
show the corresponding 1D probability distributions.

For comparison, we also construct 10,000 subsamples, each containing $19$ clusters from our MS sample.
The subsamples satisfy the same selection criteria as that shown in the bottom panels of Figure \ref{fig:eobs19},
\ie only clusters with reduced $\chi^2\le2$ for NFW fitting, observed $c\sim[1.5,8]$ and the true mass larger than $3.57M_{14}$
are included. The Bayesian fitting results for these simulated subsamples are shown in black in Figure \ref{fig:Simok19}.
It is seen that although the results of Ok10 for $(A, \alpha)$
are somewhat biased in comparison to the subsamples based on the MS simulation, the differences are within
$1-\sigma$ ranges. Therefore, again, we see that by properly modeling the noise effects and the selection function,
the higher $A$ and the steeper $\alpha$ seen in Ok10 may not pose serious challenges to the $\Lambda {\rm CDM}$ cosmology.

It should be mentioned that Ok10 point out that their sample is indistinguishable from a volume-limited sample
with X-ray luminosity $L_X/E(z)^{2.7}\ge4.2\times10^{44}{\rm erg}\,{\rm s}^{-1}$, where $E(z)=H(z)/H_0$.
Therefore, it is appropriate to regard their sample as a mass limited one and to use $M_{\rm low}$ to model the selection effect in
our Bayesian analyses.

On the other hand, for Og12, their sample is strong-lensing selected. The selection function is more complicated
than that of the mass limited one. In principle, we can incorporate the selection function into the
Bayesian analyses. On the other hand, because Og12 use the simple $\chi^2$ fitting in their $c$--$M$ relation analyses,
here we choose to only perform Monte Carlo studies for Og12 with the simple $\chi^2$ fitting. In other words,
we compare the apparent $c$--$M$ relation.

Og12 analyze $28$ clusters selected by strong lensing properties through the Sloan Giant Arcs Survey (SGAS)
from the SDSS. The average redshift of the clusters is $z\approx 0.46$.
The weak-lensing observations are done with the Subaru/Suprime-cam.
For each cluster, they analyze the density profile in two ways, using the weak-lensing data alone and
combining the weak-lensing data with the Einstein radius determined from the strong lensing giant arcs.
For the latter, they obtain the $c$--$M$ relation corresponding to $\Delta_{\rm vir}$ with $A\approx 7.7\pm 0.6$ and
$\alpha\approx -0.59\pm 0.12$ with the pivot mass ${M}_p=5\times 10^{14}h^{-1}\msun$.
Using the weak-lensing only results, shown in their Table 4 and converting them to $(c, M)$ corresponding to $\Delta=200$,
we fit the $c$--$M$ relation and get $A=5.28\pm 0.6$ and $\alpha= -0.94\pm 0.12$ for the pivot mass
${M}_p=5\times 10^{14}h^{-1}\msun$.

For the Monte Carlo studies, we use our $1690+66$ simulated halos at $z=0.46$.
For each halo, we populate source galaxies at $z=1.12$ with the number density $n_g$ chosen randomly in the range
of $[7, 18]\hbox{ arcmin}^{-2}$ and $\sigma_{\epsilon_s}=0.4$, consistent with the values in Og12.
We perform the NFW fitting to the 1D reduced tangential shear signals
to obtain $(c, M)$ for each map. The default fitting range is from $0\overset{\prime}{.}4$ to $8\arcmin$ and the concentration is limited in the range of $[0.01,39.81]$. We then randomly select
$28$ clusters with the weak-lensing-derived mass,
$\ge 0.5\times10^{14}h^{-1}\msun$, to form one realization and generate $1000$ realizations in total.
We also generate $1000$ realizations for the case without noise for comparison. To take into account the
strong lensing selection, we estimate the Einstein radius $\theta_E$ with the strong lensing source at redshift $z=2$
for each halo by looking for the place where the mean convergence $\overline{\kappa}(\theta_E)=1$.
Adding the criterion of $\theta_E >5''$, 191 low-mass and 44 high-mass halos are selected. From them we regenerate
$1000$ realizations each containing $28$ clusters. The same procedures are also performed for subsamples with a larger Einstein radius, $\theta_E >8''$.

\begin{figure*}
  \centering
  \includegraphics[width=0.81\textwidth]{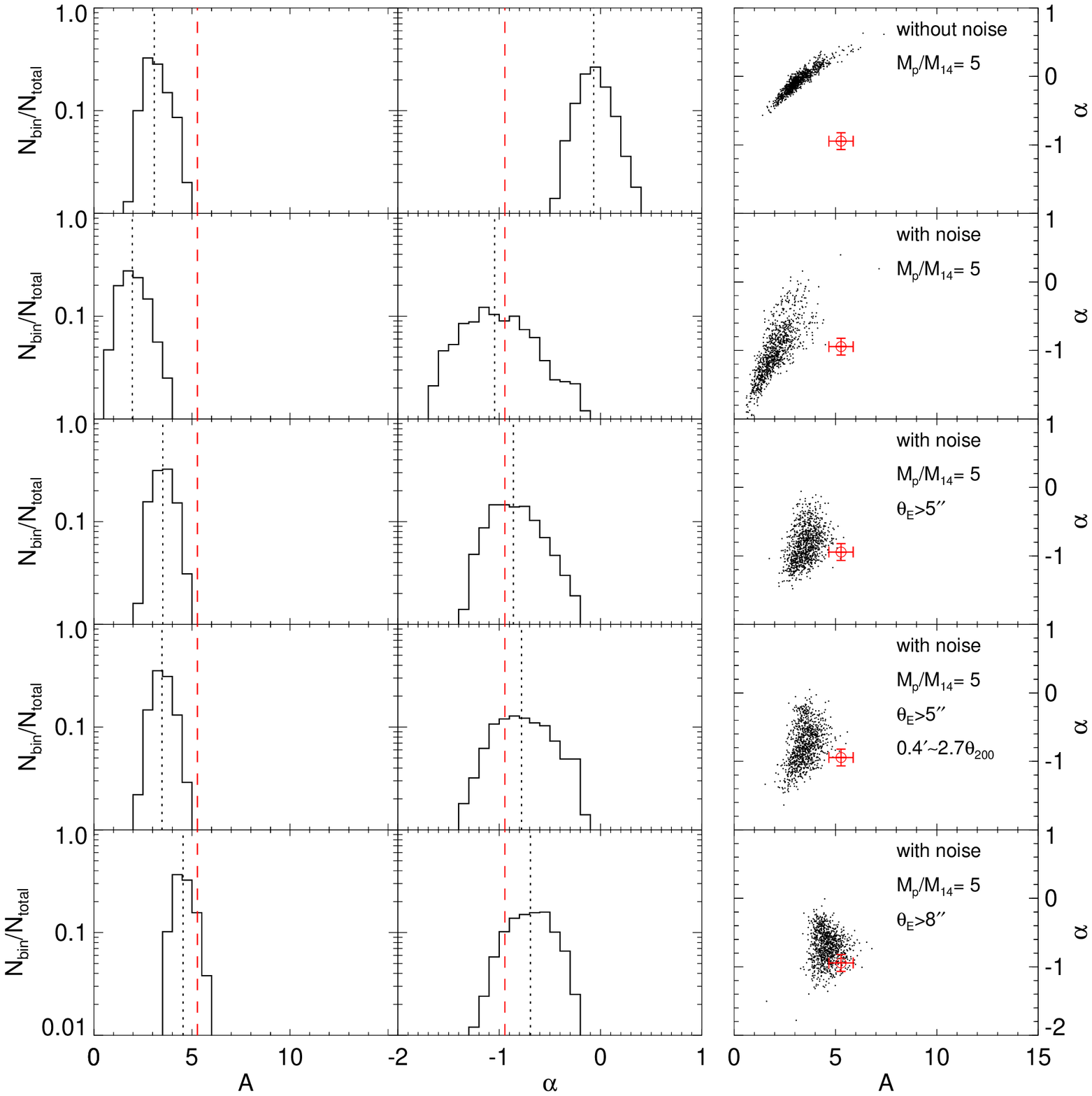}\\
  \caption{Results from $1000$ mock realizations constructed from MS clusters mimicking the $28$ clusters of Og12.
The first and second rows are for the 1000 realizations
without and with noise, respectively. The third row shows the results for strong lensing-selected samples with $\theta_{E}\ge 5\arcsec$.
The fourth row is for the results with the NFW fitting range of $[0\overset{\prime}{.}4, 2.7\theta_{200}]$.
The bottom row shows the results for samples with $\theta_{E}\ge 8\arcsec$.
The vertical red dashed lines and the red circles with error bars are for the result from Og12.
The left two columns are for the probability distributions of $A$ and $\alpha$, respectively. The vertical dotted lines show the median
values of the distributions. The right column shows the results on the $A$--$\alpha$ plane.
Here the pivot mass ${M}_p=5\times{M}_{14}$ is adopted.}\label{fig:eobs28}
\end{figure*}

The results are shown in Figure \ref{fig:eobs28},
where the red lines and symbols are for the weak-lensing only result of Og12 with $A=5.28\pm 0.6$ and $\alpha= -0.94\pm 0.12$.
In the case without noise (first row), the Og12 result is grossly away from the bulk of the Monte Carlo results,
showing a strong inconsistency. With the noise included (second row), the slope $\alpha$ is strongly enhanced,
and the Og12 result is in good agreement with that from Monte Carlo analyses. However, the value of $A$ from Og12 is
still significantly higher.

Further taking into account the strong lensing selection with $\theta_E >5''$ (third row), which tends to pick up
halos with high concentrations, the agreement between the Og12 result and the Monte Carlo results improves considerably.
Still, $A$ from Og12 is higher than that from the Monte Carlo simulations. In Og12's analysis, the fitting range
is up to about $2.7 \theta_{200}$, on average, where $\theta_{200}=r_{200}/D_d(z=0.46)$. For a test, we
regenerate $1000$ realizations by using the fitting range of $[0\overset{\prime}{.}4,2.7\theta_{200}]$
for each halo with much a larger FOV. The result is presented in the fourth row of Figure \ref{fig:eobs28}. The slope $\alpha$ from Monte Carlo analyses is slightly flatter
but the normalization is almost unchanged. The last row presents the results with $\theta_E >8''$.
With the selection of this larger Einstein radius, the agreement between the simulation results and that of Og12 improves further
but the differences are still seen and $A$ is higher in Og12.

It is noted that in both Ok10 and Og12, they discuss whether the steeper $\alpha$ can be attributed
to the known degeneracy between $c$ and $M$. They randomly sample $(c, M)$ in the region corresponding to the
posterior distributions of weak-lensing-determined $c$ and $M$ for each cluster and then perform a
simple $\chi^2$ fitting to the resampled data to derive the $c$--$M$ relation. They conclude that
the effects are not significant. It should be noted however, that the posterior distributions are
around the best-fit values of $c$ and $M$, which can already be deviated from the reference $(c, M)$
(see Figure \ref{fig:cm1556}). Therefore, such resampling cannot fully reveal the noise effects if the simple $\chi^2$ fitting is done for analyzing the $c$--$M$ relation.
In Ok10, they also perform Monte Carlo analyses based on an artificial cluster sample without intrinsic
correlations between $c$ and $M$. They find $\alpha\sim -0.06$ from Monte Carlo weak-lensing analysis with noise.
For our cluster samples, the underlying $c$--$M$ relation for dark matter halos has $\alpha\sim -0.1$.
Thus, the increase of median $|\alpha|$ due to noise (the middle panel of the second row of Figure \ref{fig:eobs19}) is $\sim 0.164-0.1=0.06$,
consistent with the result of Ok10. Although the bias of $\alpha$ is small for the noise level considered here, the scatter is significantly increased by the noise.

In summary, our Monte Carlo analyses show that by including a proper mass selection function $M_{\rm low}$,
the observational result of Ok10 is, within $1-\sigma$ range, consistent with the simulation results based on the MS cluster sample,
and the apparently steep slope $\alpha$ for the $c$--$M$ relation can be largely explained by the noise effects and the sample variance.
For Og12, the noise effects can account for the steep $\alpha$ but the amplitude $A$ of the derived $c$--$M$ relation is
considerably higher than that of simulations. The probability of finding Og12's result is somewhat low even for a selection function with a quite large Einstein radius.

\section{Discussion}

Based on the halo catalog from the {\it Millennium Simulation}, we generate mock weak-lensing data for $\sim 2000$ clusters and systematically investigate the effects of different center finding methods and the noise on the weak-lensing-derived $c$--$M$ relation.

Four different methods of center identification using weak-lensing data alone are explored. In agreement with
other studies \citep[e.g.,][]{2012MNRAS.419.3547D}, we find that both the shape noise and the smoothing can lead to misidentifications of centers.
For high-mass halos, their intrinsically irregular mass distribution can also considerably affect the center identifications.
In general, the performance of a center identification method in a weak-lensing regime mainly depends on the ${\rm S/N}$.
For intermediate redshift clusters, the quantity $E_{\rm SN}=(D_{ds}/D_s)/\sigma_n$ can be used
as an indicator to estimate the efficiency of weak-lensing center identification methods. Among different methods, our newly proposed
two-scale smoothing method, 2K05K, performs the best. Our simulation analyses show that for $E_{\rm SN}\sim10$,
at least $90\%$ of the identified centers by 2K05K have offsets to the true centers being less than $r_s$, demonstrating its great
potential in weak-lensing cluster studies.

For NFW profile constraints, consistent with other studies \citep[e.g.,][]{2010MNRAS.405.2078M, 2012ApJ...757....2G}, we notice that the center offset alone
can lead to a large negative bias to the fitted mass and concentration parameter if it is larger than $r_s$.
On the other hand, unlike studies in which centers are defined independent of weak-lensing data,
we identify centers from weak-lensing analyses. Therefore, the located centers do have high apparent weak-lensing signals,
either associated with real subclumps for massive halos or due to the chance alignments of source galaxies leading to high false peaks
for relatively low-mass halos. Consequently, with these contributions, the net biases on the mass and concentration parameter are
actually smaller than that from the pure effect of center offset without including these additional signals.

Concentrating on statistically analyzing the $c$--$M$ relation from a sample of weak-lensing studied clusters, each with an NFW fitted $(c,M)$,
we find that for our default case with $n_g=30\hbox{ arcmin}^{-2}$ and $\sigma_{\epsilon_s}=0.4$,
using centers identified by 2K05K gives rise to almost the same result on the $c$--$M$ relation
as that using the true centers in the NFW profile fitting. Again, this shows the feasibility of finding centers in weak-lensing cluster analyses.

Regarding the shape noise effects, our systematic analyses, assuming different noise levels, reveal that they can significantly affect
the apparent $c$--$M$ relation from weak-lensing studies. If the simple $\chi^2$ fitting method is adopted
in deriving the $c$--$M$ relation from a sample of weak-lensing analyzed clusters, we generally obtain
a steeper slope parameter $\alpha$, in line with a number of observational studies \citep[e.g.,][Og12, Ok10]{2013MNRAS.434..878S}.
This is mainly due to the scatter associated with the weak-lensing determined $(c,M)$ and their covariance, which
are not properly accounted for in the simple $\chi^2$ fitting.

For the stacking analyses based on the weak-lensing-derived mass, similar problems exist; thus similar noise effects occur
and lead to a steeper $c$--$M$ relation than the true one.
On the other hand, if the stacking is based on true mass, the derived $c$--$M$ relation shows no significant bias
with respect to the true $c$--$M$ relation. Observationally, we do not directly know the true mass of halos.
However, if we have other weak-lensing independent observables that strongly correlate with the underlying true mass of halos,
they can be used to group clusters for stacking. We then expect no significant bias in the derived $c$--$M$ relation.
For that, the galaxy-galaxy lensing technique can be advantageous for either the luminosity or the stellar mass of foreground galaxies, which
are often used to form bins for stacking analyses \citep[e.g.,][]{2006MNRAS.368..715M, 2009MNRAS.394.1016L}. They are shown to be good tracers of
their host halo mass \citep{2007ApJ...671..153Y}. The galaxy-galaxy lensing analyses of \citet{2008JCAP...08..006M} indeed give rise to
a $c$--$M$ relation that is rather consistent with the 3D result with a slope of $\alpha\sim -0.1$.

On the other hand, concerning weak-lensing analyses alone, and the apparent steep $c$--$M$ relation seen from the
simple $\chi^2$ fitting, we further develop a Bayesian method in probing the $c$--$M$ relation,
taking into account $(c,M)$ scatter and covariance. The sample selection effect is also
considered by invoking a lower mass limit for a mass selected sample. In principle, more complicated
selection functions suitable for different samples can be incorporated into the Bayesian method.
Our results show that with the Bayesian method, we can extract the underlying $c$--$M$ relation of dark matter halos faithfully
from the apparently noise-affected data for a wide range of noise levels.

Having finished our analyses, we note a recent study by \citet{2013MNRAS.436..503A}. They perform constraints on the $c$--$M$ relation
by combining the information of Einstein radius and optical richness for $26$ group and cluster-scale objects also using
a Bayesian approach. They also point out that the steep slope found by some previous studies is likely due to the degeneracy between concentration and mass.

To account for the sample variance of samples containing a limited number of clusters, we conduct Monte Carlo analyses with respect to
the samples of Ok10 and Og12. We find that the noise effects significantly
contribute to the apparently steep $\alpha$ in the $c$--$M$ relation obtained by Ok10 and by Og12.
Our Bayesian analyses show that if a lower mass limit $M_{\rm low}=3.57\times10^{14}h^{-1}\msun$ is adopted,
the Ok10 result is reasonably consistent with that of the MS clusters within $1-\sigma$ range.
For Og12, while the slope $\alpha$ is consistent with that of MS clusters including the noise effects,
the amplitude parameter $A$ in the $c$--$M$ relation is higher from Og12, even with the strong-lensing bias
taken into consideration.

It is noted that MS is a dark matter-only simulation without including baryonic physics.
It is believed that baryonic cooling and star formation tend to increase the concentration of the halo mass distribution \citep[e.g.,][]{2010MNRAS.405.2161D,2012MNRAS.426.1558G,2011MNRAS.416.2539K}.
On the other hand, feedback from supernova and AGNs can heat the gas and suppress star formation. Therefore,
the overall effects of the baryonic component on the density profile of cluster-scale halos may not be significant \citep[e.g.,][]{2010MNRAS.405.2161D,2010MNRAS.406..434M,2012MNRAS.424.1244F}.

It should also be pointed out that MS adopts the cosmological model with a higher $\sigma_8=0.9$ and lower $\Omega_\smM=0.25$
than the current best model from latest observations, \eg {\it WMAP}7 and {\it Planck}  \citep{2013arXiv1303.5076P,2011ApJS..192...18K}.
Varying cosmological models can affect the properties of dark matter halos and therefore their $c$--$M$ relation \citep{2003ApJ...598...49Z,2011MNRAS.418.2422R}.
It is shown that while the slope $\alpha$ is not sensitive to cosmological models, the amplitude $A$ is \citep[e.g.,][]{2004A&A...416..853D,2008MNRAS.391.1940M,2008MNRAS.390L..64D}.
For instance, $A$ for {\it WMAP}1 cosmology is about $19\%$ higher than that of {\it WMAP}5 model, and the slope $\alpha$ is consistent
with $-0.1$ in both models \citep{2008MNRAS.391.1940M,2008MNRAS.390L..64D}.

However, very recent studies perform a direct comparison of halo concentration from the MS and predict from the {\it Planck} cosmological model \citep{2014MNRAS.441..378L}.
The results show that for cluster-scale halos, the concentration parameter is nearly the same for the two models.
This is because the {\it Planck} cosmological model has a larger $\Omega_m$ than that of MS, although $\sigma_8$ is smaller.

Our analyses for massive halos show that substructures can affect the center identifications, and also lead to a negative bias to
2D concentration in comparison with that of 3-D halos \citep[see also][]{2012MNRAS.426.1558G,2012MNRAS.421.1073B}.
It has been shown that a halo with lower concentration and higher mass and a halo formed later tend to contain a higher fraction of mass in substructures
\citep{2004MNRAS.355..819G,2010MNRAS.404..502G, 2012MNRAS.420.2978C}. The change of cosmology is expected to affect the mass accretion history and therefore the subhalo abundances.
However, recent analyses of \citet{2012MNRAS.425.2169G} and \citet{2012MNRAS.424.2715W} show that substructure abundances are not very sensitive to cosmological models.
For example, \citet{2012MNRAS.424.2715W} find that subhalo abundances at given host halo mass are very similar between MS and {\it WMAP}7 cosmology,
although MS has a higher $\sigma_8$.

Overall, our results presented here based on MS-simulated clusters may not be seriously affected by the
specific cosmological model adopted by MS. The reasonable agreement between our simulation analyses, including the
noise and selection effects and that of Ok10 for the $c$--$M$ relation, may indicate that our current understanding of the
structure formation is well on track. On the other hand, the somewhat larger $A$ seen in Og12 still deserves further
studies.

Apart from comparing with Ok10 and Og12, in this paper, we emphasize the methodology of center identification using weak-lensing analyses alone,
and the noise effects on the weak-lensing-derived $c$--$M$ relation. Future weak-lensing observations can result in a large number of
clusters with high signal-to-noise ratios. We expect that both our newly proposed 2K05K center identification method, and the
Bayesian method in deriving the $c$--$M$ relation from a sample of weak-lensing studied clusters can have important applications
in future cosmological studies.

\acknowledgments
We thank the referee for the constructive and detailed suggestions that have resulted in significant improvements of our studies.
We are very grateful to Liang Gao and Ran Li for providing halo catalogs from the {\it Millennium Simulation},
which was carried out as part of the Virgo Consortium. We also thank Toshifumi Futamase for stimulating discussions.
This research is supported in part by the NSFC of China under grants 11333001, 11173001, and 11033005, and the 973 program No. 2007CB815401.

\appendix
\section{Bayesian approach in probing $\lowercase{c}$--$M$ relation}

In the Bayesian analyses shown in Section 4.2.2, the distribution of $P(c_{\rm ob},M_{\rm ob}|c_\smT,M_\smT,\sn)$ is crucial. Modeled as a 2D Gaussian
distribution, it can be written as

\begin{equation}\label{eq:Apcm_onT}
P(c_{\rm ob},M_{\rm ob}|c_\smT,M_\smT,\sn)=\frac{1}{2\pi\sigma_\smM\sigma_c\sqrt{1-r_{c\smM}^2}}\exp\left[-\frac{\sigma_c^2\lgg^2\frac{M_{\rm ob}}{M_\smT}+\sigma_\smM^2
\lgg^2\frac{c_{\rm ob}}{c_\smT}-2r_{c\smM}\sigma_\smM\sigma_c \lgg\,\frac{M_{\rm ob}}{M_\smT}\lgg\,\frac{c_{\rm ob}}{c_\smT}}{2(1-r_{c\smM}^2)\sigma_\smM^2\sigma_c^2}\right],
\end{equation}
where the inverse matrix $\mathcal{C}^{-1}$ in Equation \ref{eq:pcm_onT} is written out explicitly.

To analyze the dependencies of $\sigma_{\smM}$, $\sigma_c$, and $r_{c\smM}$ on the noise level $\sigma_n$ and the mass and concentration parameter
of halos, we consider $18$ different noise levels with $\sigma_n$ in the range of $[0.03, 0.2]$ for each of the $1756$ halos of our parent sample.
At each $\sigma_n$, we generate $100$ mock weak-lensing data with different realizations of the source galaxy distribution and
intrinsic ellipticities for each cluster, and perform NFW fitting to each of the mock data. From the $100$ best fit $(c,M)$ of a cluster with
the true $(c_T, M_T)$, we can then calculate the corresponding $\sigma_{\smM}$, $\sigma_c$ and $r_{c\smM}$. We find that they are
sensitive to $\sigma_n$ and $M_T$ (see Figure \ref{fig:cm1556} for example) but nearly independent of $c_T$.

For our specific catalogs at $z_d=0.2$, when the NFW fitting range is from $1\arcmin$ to $15\arcmin$, we obtain

$ $\\
$\sigma_\smM(M_\smT,\sn)=\left[-0.71\snsq+2.73\sn-0.02\right](\frac{M_\smT}{M_{14}})^{7.48\snsq-1.27\sn-0.46}$,\\\\
$\sigma_c(M_\smT,\sn)=\left[(6.23\sn)^4-(9.77\sn)^3+(12.66\sn)^2\right](\frac{M_\smT}{M_{14}})^{6.83(\sn-0.09)^{1.10}-1}$,\\\\
$r_{c\smM}(M_\smT,\sn)=\left[0.94{\rm exp}\left[-\frac{(\sn-0.2)^2}{0.09^2}\right]+0.40\right](\frac{M_\smT}{M_{14}})^{0.57{\rm exp}\left[-\frac{(\sn-0.045)^2}{0.06^2}\right]-0.81}-1.$\\\\

For the fitting range from $1\arcmin$ to $10\arcmin$, we have

$ $\\
$\sigma_\smM(M_\smT,\sn)=\left[-6.45\snsq+4.19\sn-0.05\right](\frac{M_\smT}{M_{14}})^{9.98\snsq-0.66\sn-0.43},$\\\\
$\sigma_c(M_\smT,\sn)=\left[(6.98\sn)^4-(10.93\sn)^3+(14.20\sn)^2\right](\frac{M_\smT}{M_{14}})^{3.60(\sn-0.075)^{0.76}-1},$\\\\
$r_{c\smM}(M_\smT,\sn)=\left[1.04{\rm exp}\left[-\frac{(\sn-0.2)^2}{0.07^2}\right]+0.30\right](\frac{M_\smT}{M_{14}})^{0.83 {\rm exp}\left[-\frac{(\sn-0.045)^2}{0.07^2}\right]-1.13}-1.$\\\\

For the intrinsic distribution of the concentration parameter of $c_\smT$ given $M_\smT$, we use
\begin{equation}\label{eq:ctmt}
\left.
\begin{array}{ll}
P(c_\smT|M_\smT){\it d}c_\smT=\displaystyle\frac{1}{\sqrt{2\pi}\sigma_{\rm in}}\exp\left[-\frac{(\lgg\,c_\smT- \lgg\,\langle c_\smT\rangle)^2}{2\sigma_{\rm in}^2}\right]{\it d}\lgg\,c_\smT,
\end{array}
\right.
\end{equation}
where $\sigma_{\rm in}$ is the intrinsic dispersion of $c_\smT$. We take $\sigma_{\rm in}\simeq0.12$ in our calculations. We have tested $\sigma_{\rm in}\simeq0.17$ in accord with the dispersion of the 2D reference case (see the upper right panel of Figure \ref{fig:cm_relation}), and
the results change little.\\

With $P(c_{\rm ob},M_{\rm ob}|c_\smT,M_\smT,\sn)$ and $P(c_\smT|M_\smT)$, we can further obtain
the probability of the observed $(c_{\rm ob},M_{\rm ob})$ for a halo with true mass $M_\smT$, which is given by

\begin{equation}\label{eq:Apcm_onMT}
P(c_{\rm ob},M_{\rm ob}|M_\smT,\sn)=\int P(c_{\rm ob},M_{\rm ob}|c_\smT,M_\smT,\sn)P(c_\smT|M_\smT) {\it d}c_\smT.
\end{equation}

With Equations (\ref{eq:Apcm_onT}) and (\ref{eq:ctmt}), $P(c_{\rm ob},M_{\rm ob}|M_\smT,\sn)$ can be expressed by
\begin{equation}
P(c_{\rm ob},M_{\rm ob}|M_\smT,\sn)=
\frac{1}{2\pi\tilde{\sigma}}{\rm exp}\big\{-\frac{(\sigma_{\rm in}^2+\sigma_c^2)\lgg^2\frac{M_{\rm ob}}{M_T}+\sigma_M^2
\lgg^2\frac{c_{\rm ob}}{\langle c_T\rangle}-2r_{cM}\sigma_M\sigma_c \lgg\,\frac{M_{\rm ob}}{M_T}\lgg\,\frac{c_{\rm ob}}{\langle c_T\rangle}}{2\tilde{\sigma}^2}\big\}, 
\end{equation}
with $\tilde{\sigma}^2=\sigma_\smM^2\left[\sigma_{\rm in}^2+\sigma_c^2-r_{cM}^2\sigma_c^2\right]$ and $\lgg\,\langle c_\smT\rangle=\langle \lgg\,c_\smT\rangle=\lgg\,A+\alpha \lgg\,\frac{M_\smT}{M_p}$, where $A$ and $\alpha$ are the normalization and slope of the $c$--$M$ relation (see Equation (\ref{eq:cm})).\\

Considering a mass-selected sample of clusters, we then have the probability of $(c_{\rm ob},M_{\rm ob})$ given $\sn$ as

\begin{equation}
P(c_{\rm ob},M_{\rm ob}|\sn)=\frac{\int_{{M}_{\rm low}}^{\infty} P(c_{\rm ob},M_{\rm ob}|M_\smT,\sn){\it d}{\rm n}(M_\smT)}{\int_{{M}_{\rm low}}^{\infty}{\it d}{\rm n}(M_\smT)},
\end{equation}
and the probability of $c_{\rm ob}$ given $(M_{\rm ob},\sn)$ as

\begin{equation}
P(c_{\rm ob}|M_{\rm ob},\sn)=\frac{\int_{{M}_{\rm low}}^{\infty} P(c_{\rm ob},M_{\rm ob}|M_\smT,\sn){\it d}{\rm n}(M_\smT)}{\int_{{M}_{\rm low}}^{\infty}
P(M_{\rm ob}|M_\smT,\sn){\it d}{\rm n}(M_\smT)},
\end{equation}
where $M_{\rm low}$ denotes the lower mass limit of the sample and ${\rm n}(M_\smT)$ is the halo mass function. The probability of $M_{\rm ob}$ for a halo with mass $M_\smT$ can be written as
\begin{equation}
P(M_{\rm ob}|M_\smT,\sn){\it d}M_{\rm ob}=\frac{1}{\sqrt{2\pi}\sigma_\smM}exp(-\frac{(\lgg M_{\rm ob}-\lgg M_\smT)^2}{2\sigma_\smM^2}){\it d}\lgg M_{\rm ob}. 
\end{equation}

For the mass function ${\rm n}(M_\smT,z)$ in comoving coordinates, we adopt the Tinker form \citep{2008ApJ...688..709T} given by
\begin{equation}
{\it d}{\rm n}(M_\smT,z)=f(\sigma)\frac{\Omega_m\rho_{\rm crit,0}}{M_\smT}\frac{{\it d}{\rm ln}\sigma^{-1}}{{\it d}M_\smT}{\it d}M_\smT,
\end{equation}
where $\rho_{\rm crit,0}$ is the current critical density of the universe, $\sigma(M_\smT,z)$ is the rms of matter perturbations at redshift $z$ and filtered on a scale enclosing mass $M_\smT$.
The fitting formula for $f$ is
\begin{equation}
f(\sigma)=A_{\rm \smT in}[(\sigma/b_{\rm \smT in})^{-a_{\rm \smT in}}+1]{\rm exp}(-c_{\rm \smT in}/\sigma^2),
\end{equation}
where $A_{\rm \smT in}$, $a_{\rm \smT in}$, $b_{\rm \smT in}$ and $c_{\rm \smT in}$ are the best-fit parameters which depend on redshift and overdensity $\Delta$. For our mass definition with $200\rho_{\rm crit}$, we get $A_{\rm \smT in}=0.218$, $a_{\rm \smT in}=1.701$, $b_{\rm \smT in}=1.777$ and $c_{\rm \smT in}=1.420$ at $z=0.2$.

For large samples, the expected median concentration $\langle c_{\rm ob,model}\rangle$ within a mass bin based on $M_{\rm ob}$ can be determined by
\begin{equation}
\int_{\langle c_{\rm ob,model}\rangle}^{\infty}P(c_{\rm ob}|M_{\rm ob},\sn){\rm d}c_{\rm ob}=\frac{1}{2},
\end{equation}
and they are fitted to the corresponding observational data to derive the underlying $c$--$M$ relation.

For small samples, the $c$--$M$ relation is estimated by maximizing the likelihood
\begin{equation}
\mathcal{L}=\displaystyle \prod_i P_i(c_{\rm ob},M_{\rm ob}|\sn).
\end{equation}


\begin{thebibliography}{}
\bibitem[Auger et al.(2013)]{2013MNRAS.436..503A} Auger, M.~W., Budzynski,J.~M., Belokurov, V., Koposov, S.~E.,\& McCarthy, I.~G.\ 2013, \mnras, 436, 503\\[-20pt]
\bibitem[Bah{\'e} et al.(2012)]{2012MNRAS.421.1073B} Bah{\'e}, Y.~M.,McCarthy, I.~G., \& King, L.~J.\ 2012, \mnras, 421, 1073\\[-20pt]
\bibitem[Bartelmann\& Schneider(2001)]{2001PhR...340..291B} Bartelmann, M., \& Schneider, P.\ 2001, \physrep, 340, 291\\[-20pt]
\bibitem[Becker\& Kravtsov(2011)]{2011ApJ...740...25B} Becker, M.~R., \& Kravtsov, A.~V.\ 2011, \apj, 740, 25\\[-20pt]
\bibitem[Bhattacharya et al.(2013)]{2013ApJ...766...32B} Bhattacharya, S.,Habib, S., Heitmann, K., \& Vikhlinin, A.\ 2013, \apj, 766, 32\\[-20pt]
\bibitem[Bryan\& Norman(1998)]{1998ApJ...495...80B} Bryan, G.~L., \& Norman, M.~L.\ 1998, \apj, 495, 80\\[-20pt]
\bibitem[Bullock et al.(2001)]{2001MNRAS.321..559B} Bullock, J.~S., Kolatt,T.~S., Sigad, Y., et al.\ 2001, \mnras, 321, 559\\[-20pt]
\bibitem[Clowe et al.(2004)]{2004MNRAS.350.1038C} Clowe, D., De Lucia, G.,\& King, L.\ 2004, \mnras, 350, 1038\\[-20pt]
\bibitem[Clowe etal.(2006)]{2006A&A...451..395C} Clowe, D., Schneider, P., Arag{\'o}n-Salamanca, A., et al.\ 2006, \aap, 451, 395\\[-20pt]
\bibitem[Comerford\& Natarajan(2007)]{2007MNRAS.379..190C} Comerford, J.~M., \& Natarajan, P.\ 2007, \mnras, 379, 190\\[-20pt]
\bibitem[Contini et al.(2012)]{2012MNRAS.420.2978C} Contini, E., De Lucia,G., \& Borgani, S.\ 2012, \mnras, 420, 2978\\[-20pt]
\bibitem[Corless\& King(2007)]{2007MNRAS.380..149C} Corless, V.~L., \& King, L.~J.\ 2007, \mnras, 380, 149\\[-20pt]
\bibitem[Davis et al.(1985)]{1985ApJ...292..371D} Davis, M., Efstathiou,G., Frenk, C.~S., \& White, S.~D.~M.\ 1985, \apj, 292, 371\\[-20pt]
\bibitem[De Boni et al.(2013)]{2013MNRAS.428.2921D} De Boni, C., Ettori,S., Dolag, K., \& Moscardini, L.\ 2013, \mnras, 428, 2921\\[-20pt]
\bibitem[Dietrich et al.(2012)]{2012MNRAS.419.3547D} Dietrich, J.~P.,B{\"o}hnert, A., Lombardi, M., Hilbert, S.,\& Hartlap, J.\ 2012, \mnras, 419, 3547\\[-20pt]
\bibitem[Dodelson(2004)]{2004PhRvD..70b3008D} Dodelson, S.\ 2004, \prd, 70,023008\\[-20pt]
\bibitem[Dolag etal.(2004)]{2004A&A...416..853D} Dolag, K., Bartelmann, M., Perrotta, F., et al.\ 2004, \aap, 416, 853\\[-20pt]
\bibitem[Duffy et al.(2008)]{2008MNRAS.390L..64D} Duffy, A.~R., Schaye, J.,Kay, S.~T., \& Dalla Vecchia, C.\ 2008, \mnras, 390, L64\\[-20pt]
\bibitem[Duffy et al.(2010)]{2010MNRAS.405.2161D} Duffy, A.~R., Schaye, J.,Kay, S.~T., et al.\ 2010, \mnras, 405, 2161\\[-20pt]
\bibitem[Einasto(1965)]{1965TrAlm...5...87E} Einasto, J.\ 1965, TrudyAstrofizicheskogo Instituta Alma-Ata, 5, 87\\[-20pt]
\bibitem[Ettori etal.(2010)]{2010A&A...524A..68E} Ettori, S., Gastaldello, F., Leccardi, A., et al.\ 2010, \aap, 524, A68\\[-20pt]
\bibitem[Fan et al.(2010)]{2010ApJ...719.1408F} Fan, Z., Shan, H.,\& Liu, J.\ 2010, \apj, 719, 1408\\[-20pt]
\bibitem[Fedeli(2012)]{2012MNRAS.424.1244F} Fedeli, C.\ 2012, \mnras, 424,1244\\[-20pt]
\bibitem[Gao et al.(2004)]{2004MNRAS.355..819G} Gao, L., White, S.~D.~M.,Jenkins, A., Stoehr, F., \& Springel, V.\ 2004, \mnras, 355, 819\\[-20pt]
\bibitem[Gao et al.(2008)]{2008MNRAS.387..536G} Gao, L., Navarro, J.~F.,Cole, S., et al.\ 2008, \mnras, 387, 536\\[-20pt]
\bibitem[Gao et al.(2012)]{2012MNRAS.425.2169G} Gao, L., Navarro, J.~F.,Frenk, C.~S., et al.\ 2012, \mnras, 425, 2169\\[-20pt]
\bibitem[George et al.(2012)]{2012ApJ...757....2G} George, M.~R.,Leauthaud, A., Bundy, K., et al.\ 2012, \apj, 757, 2\\[-20pt]
\bibitem[Giocoli et al.(2010)]{2010MNRAS.404..502G} Giocoli, C., Tormen,G., Sheth, R.~K., \& van den Bosch, F.~C.\ 2010, \mnras, 404, 502\\[-20pt]
\bibitem[Giocoli et al.(2012)]{2012MNRAS.426.1558G} Giocoli, C.,Meneghetti, M., Ettori, S., \& Moscardini, L.\ 2012, \mnras, 426, 1558\\[-20pt]
\bibitem[Giocoli et al.(2014)]{2014MNRAS.440.1899G} Giocoli, C.,Meneghetti, M., Metcalf, R.~B., Ettori, S.,\& Moscardini, L.\ 2014, \mnras, 440, 1899\\[-20pt]
\bibitem[Hamana et al.(2004)]{2004MNRAS.350..893H} Hamana, T., Takada, M.,\& Yoshida, N.\ 2004, \mnras, 350, 893\\[-20pt]
\bibitem[Hernquist(1990)]{1990ApJ...356..359H} Hernquist, L.\ 1990, \apj,356, 359\\[-20pt]
\bibitem[Hockney\& Eastwood(1981)]{1981csup.book.....H} Hockney, R.~W., \& Eastwood, J.~W.\ 1981, Computer Simulation Using Particles, New York: McGraw-Hill, 1981,\\[-20pt]
\bibitem[Hoekstra et al.(2000)]{2000ApJ...532...88H} Hoekstra, H., Franx,M., \& Kuijken, K.\ 2000, \apj, 532, 88\\[-20pt]
\bibitem[Hoekstra et al.(2011)]{2011MNRAS.412.2095H} Hoekstra, H., Hartlap,J., Hilbert, S., \& van Uitert, E.\ 2011, \mnras, 412, 2095\\[-20pt]
\bibitem[Hoekstra et al.(2013)]{2013SSRv..177...75H} Hoekstra, H.,Bartelmann, M., Dahle, H., et al.\ 2013, \ssr, 177, 75\\[-20pt]
\bibitem[Hoekstra(2003)]{2003MNRAS.339.1155H} Hoekstra, H.\ 2003, \mnras,339, 1155\\[-20pt]
\bibitem[Hu\& Kravtsov(2003)]{2003ApJ...584..702H} Hu, W., \& Kravtsov, A.~V.\ 2003, \apj, 584, 702\\[-20pt]
\bibitem[Israel etal.(2010)]{2010A&A...520A..58I} Israel, H., Erben, T., Reiprich, T.~H., et al.\ 2010, \aap, 520, A58\\[-20pt]
\bibitem[Israel etal.(2012)]{2012A&A...546A..79I} Israel, H., Erben, T., Reiprich, T.~H., et al.\ 2012, \aap, 546, A79\\[-20pt]
\bibitem[Johnston et al.(2007)]{2007arXiv0709.1159J} Johnston, D.~E.,Sheldon, E.~S., Wechsler, R.~H., et al.\ 2007, arXiv:0709.1159\\[-20pt]
\bibitem[Kaiser\& Squires(1993)]{1993ApJ...404..441K} Kaiser, N., \& Squires, G.\ 1993, \apj, 404, 441\\[-20pt]
\bibitem[King\& Mead(2011)]{2011MNRAS.416.2539K} King, L.~J., \& Mead, J.~M.~G.\ 2011, \mnras, 416, 2539\\[-20pt]
\bibitem[Kneib\& Natarajan(2011)]{2011A&ARv..19...47K} Kneib, J.-P., \& Natarajan, P.\ 2011, \aapr, 19, 47\\[-20pt]
\bibitem[Koester et al.(2007a)]{2007ApJ...660..221K} Koester, B.~P., McKay,T.~A., Annis, J., et al.\ 2007, \apj, 660, 221\\[-20pt]
\bibitem[Koester et al.(2007b)]{2007ApJ...660..239K} Koester, B.~P., McKay,T.~A., Annis, J., et al.\ 2007, \apj, 660, 239\\[-20pt]
\bibitem[Komatsu et al.(2011)]{2011ApJS..192...18K} Komatsu, E., Smith,K.~M., Dunkley, J., et al.\ 2011, \apjs, 192, 18\\[-20pt]
\bibitem[Levenberg (1944)]{Levenberg1944}Levenberg K. 1944, The Quarterly of Applied Mathematics, 2, 164\\[-20pt]
\bibitem[Li et al.(2009)]{2009MNRAS.394.1016L} Li, R., Mo, H.~J., Fan, Z.,et al.\ 2009, \mnras, 394, 1016\\[-20pt]
\bibitem[Ludlow et al.(2014)]{2014MNRAS.441..378L} Ludlow, A.~D., Navarro,J.~F., Angulo, R.~E., et al.\ 2014, \mnras, 441, 378\\[-20pt]
\bibitem[Macci{\`o} et al.(2008)]{2008MNRAS.391.1940M} Macci{\`o}, A.~V.,Dutton, A.~A., \& van den Bosch, F.~C.\ 2008, \mnras, 391, 1940\\[-20pt]
\bibitem[Mandelbaum et al.(2006)]{2006MNRAS.368..715M} Mandelbaum, R.,Seljak, U., Kauffmann, G., Hirata, C.~M.,\& Brinkmann, J.\ 2006, \mnras, 368, 715\\[-20pt]
\bibitem[Mandelbaum et al.(2008)]{2008JCAP...08..006M} Mandelbaum, R.,Seljak, U., \& Hirata, C.~M.\ 2008, JCAP, 8, 6\\[-20pt]
\bibitem[Mandelbaum et al.(2010)]{2010MNRAS.405.2078M} Mandelbaum, R.,Seljak, U., Baldauf, T., \& Smith, R.~E.\ 2010, \mnras, 405, 2078\\[-20pt]
\bibitem[Mann\& Ebeling(2012)]{2012MNRAS.420.2120M} Mann, A.~W., \& Ebeling, H.\ 2012, \mnras, 420, 2120\\[-20pt]
\bibitem[Markwardt(2009)]{2009ASPC..411..251M} Markwardt, C.~B.\ 2009,Astronomical Data Analysis Software and Systems XVIII, 411, 251\\[-20pt]
\bibitem[Marquardt (1963)]{Marquardt1963}Marquardt D., 1963, SIAM Journal on Applied Mathematics, 11, 431\\[-20pt]
\bibitem[Mead et al.(2010)]{2010MNRAS.406..434M} Mead, J.~M.~G., King,L.~J., Sijacki, D., et al.\ 2010, \mnras, 406, 434\\[-20pt]
\bibitem[Mu{\~n}oz-Cuartas et al.(2011)]{2011MNRAS.411..584M}Mu{\~n}oz-Cuartas, J.~C., Macci{\`o}, A.~V., Gottl{\"o}ber, S.,\& Dutton, A.~A.\ 2011, \mnras, 411, 584\\[-20pt]
\bibitem[Navarro et al.(1996)]{1996ApJ...462..563N} Navarro, J.~F., Frenk,C.~S., \& White, S.~D.~M.\ 1996, \apj, 462, 563\\[-20pt]
\bibitem[Navarro et al.(1997)]{1997ApJ...490..493N} Navarro, J.~F., Frenk,C.~S., \& White, S.~D.~M.\ 1997, \apj, 490, 493\\[-20pt]
\bibitem[Neto et al.(2007)]{2007MNRAS.381.1450N} Neto, A.~F., Gao, L.,Bett, P., et al.\ 2007, \mnras, 381, 1450\\[-20pt]
\bibitem[Oguri et al.(2010)]{2010MNRAS.405.2215O} Oguri, M., Takada, M.,Okabe, N., \& Smith, G.~P.\ 2010, \mnras, 405, 2215\\[-20pt]
\bibitem[Oguri et al.(2012)]{2012MNRAS.420.3213O} Oguri, M., Bayliss,M.~B., Dahle, H., et al.\ 2012, \mnras, 420, 3213\\[-20pt]
\bibitem[Oguri\& Hamana(2011)]{2011MNRAS.414.1851O} Oguri, M., \& Hamana, T.\ 2011, \mnras, 414, 1851\\[-20pt]
\bibitem[Okabe et al.(2010)]{2010PASJ...62..811O} Okabe, N., Takada, M.,Umetsu, K., Futamase, T., \& Smith, G.~P.\ 2010, \pasj, 62, 811\\[-20pt]
\bibitem[Park\& Schowengerdt(1983)]{1983CGIP...23..258P} Park, S.~K., \& Schowengerdt, R.~A.\ 1983, Computer Graphics Image Processing, 23, 258\\[-20pt]
\bibitem[Planck Collaboration et al.(2013)]{2013arXiv1303.5076P} PlanckCollaboration, Ade, P.~A.~R., Aghanim, N., et al.\ 2013, arXiv:1303.5076\\[-20pt]
\bibitem[Prada et al.(2012)]{2012MNRAS.423.3018P} Prada, F., Klypin, A.~A.,Cuesta, A.~J., Betancort-Rijo, J.~E.,\& Primack, J.\ 2012, \mnras, 423, 3018\\[-20pt]
\bibitem[Retana-Montenegro etal.(2012)]{2012A&A...540A..70R} Retana-Montenegro, E., van Hese, E., Gentile, G., Baes, M., \& Frutos-Alfaro, F.\ 2012, \aap, 540, A70\\[-20pt]
\bibitem[Ruiz et al.(2011)]{2011MNRAS.418.2422R} Ruiz, A.~N., Padilla,N.~D., Dom{\'{\i}}nguez, M.~J., \& Cora, S.~A.\ 2011, \mnras, 418, 2422\\[-20pt]
\bibitem[Schirmer etal.(2007)]{2007A&A...462..875S} Schirmer, M., Erben, T., Hetterscheidt, M., \& Schneider, P.\ 2007, \aap, 462, 875\\[-20pt]
\bibitem[Schneider \etal(2006)]{Schneider2006}Schneider P., Kochanek C. \& Wambsganss J. 2006, Gravitational Lensing: Strong, Weak and Micro, Springer-Verlag Berlin Heidelberg\\[-20pt]
\bibitem[Schneider etal.(2000)]{2000A&A...353...41S} Schneider, P., King, L., \& Erben, T.\ 2000, \aap, 353, 41\\[-20pt]
\bibitem[Schneider(1996)]{1996MNRAS.283..837S} Schneider, P.\ 1996, \mnras,283, 837\\[-20pt]
\bibitem[Seitz\& Schneider(1995)]{1995A&A...297..287S} Seitz, C., \& Schneider, P.\ 1995, \aap, 297, 287\\[-20pt]
\bibitem[Seitz\& Schneider(1997)]{1997A&A...318..687S} Seitz, C., \& Schneider, P.\ 1997, \aap, 318, 687\\[-20pt]
\bibitem[Sereno\& Covone(2013)]{2013MNRAS.434..878S} Sereno, M., \& Covone, G.\ 2013, \mnras, 434, 878\\[-20pt]
\bibitem[Shan et al.(2012)]{2012ApJ...748...56S} Shan, H., Kneib, J.-P.,Tao, C., et al.\ 2012, \apj, 748, 56\\[-20pt]
\bibitem[Shaw et al.(2006)]{2006ApJ...646..815S} Shaw, L.~D., Weller, J.,Ostriker, J.~P., \& Bode, P.\ 2006, \apj, 646, 815\\[-20pt]
\bibitem[Spinelli et al.(2012)]{2012MNRAS.420.1384S} Spinelli, P.~F.,Seitz, S., Lerchster, M., Brimioulle, F.,\& Finoguenov, A.\ 2012, \mnras, 420, 1384\\[-20pt]
\bibitem[Springel et al.(2005)]{2005Natur.435..629S} Springel, V., White,S.~D.~M., Jenkins, A., et al.\ 2005, \nat, 435, 629\\[-20pt]
\bibitem[Tinker et al.(2008)]{2008ApJ...688..709T} Tinker, J., Kravtsov,A.~V., Klypin, A., et al.\ 2008, \apj, 688, 709\\[-20pt]
\bibitem[van Waerbeke(2000)]{2000MNRAS.313..524V} van Waerbeke, L.\ 2000,\mnras, 313, 524\\[-20pt]
\bibitem[Wang et al.(2012)]{2012MNRAS.424.2715W} Wang, J., Frenk, C.~S.,Navarro, J.~F., Gao, L., \& Sawala, T.\ 2012, \mnras, 424, 2715\\[-20pt]
\bibitem[Weinberg\& Kamionkowski(2002)]{2002MNRAS.337.1269W} Weinberg, N.~N., \& Kamionkowski, M.\ 2002, \mnras, 337, 1269\\[-20pt]
\bibitem[Wiesner et al.(2012)]{2012ApJ...761....1W} Wiesner, M.~P., Lin,H., Allam, S.~S., et al.\ 2012, \apj, 761, 1\\[-20pt]
\bibitem[Wright\& Brainerd(2000)]{2000ApJ...534...34W} Wright, C.~O., \& Brainerd, T.~G.\ 2000, \apj, 534, 34\\[-20pt]
\bibitem[Yang et al.(2007)]{2007ApJ...671..153Y} Yang, X., Mo, H.~J., vanden Bosch, F.~C., et al.\ 2007, \apj, 671, 153\\[-20pt]
\bibitem[York et al.(2000)]{2000AJ....120.1579Y} York, D.~G., Adelman, J.,Anderson, J.~E., Jr., et al.\ 2000, \aj, 120, 1579\\[-20pt]
\bibitem[Zentner\& Bullock(2003)]{2003ApJ...598...49Z} Zentner, A.~R., \& Bullock, J.~S.\ 2003, \apj, 598, 49\\[-20pt]
\bibitem[Zhao et al.(2009)]{2009ApJ...707..354Z} Zhao, D.~H., Jing, Y.~P.,Mo, H.~J., B{\"o}rner, G.\ 2009, \apj, 707, 354\\[-20pt]
\bibitem[Zhao(1996)]{1996MNRAS.278..488Z} Zhao, H.\ 1996, \mnras, 278, 488\\[-20pt]
\bibitem[Zitrin et al.(2012)]{2012MNRAS.426.2944Z} Zitrin, A., Bartelmann,M., Umetsu, K., Oguri, M., \& Broadhurst, T.\ 2012, \mnras, 426, 2944
\end{thebibliography}
\end{document}